# Predicting Donor Kidney Discard Using Machine Learning: Model Development and Comparison Study

Peer Schliephacke[1†], Hannah Schult[1†], Leon Mizera[1†], Judith Würfel[1†], Gunter Grieser[2], Axel Rahmel[3], Carl-Ludwig Fischer-Fröhlich[3], Antje Jahn-Eimermacher[1*]

[1]Department of Mathematics and Natural Sciences, Darmstadt University of Applied Sciences, Schöfferstrase 3, Darmstadt, 64295, Hessen, Germany.

[2]Department of Computer Science, Darmstadt University of Applied Sciences, Schöfferstrase 3, Darmstadt, 64295, Hessen, Germany.

[3]Deutsche Stiftung Organtransplantation, Deutschherrnufer 52, Frankfurt, 60594, Hessen, Germany.

*Corresponding author(s). E-mail(s): antje.jahn@h-da.de;
†These authors contributed equally to this work.

## Abstract

**Background:** A kidney transplant can significantly improve the life expectancy and quality of life of patients with end-stage renal failure. Even more patients could be helped with a kidney transplant if the rate of kidneys that are discarded and not transplanted could be reduced. Machine learning (ML) techniques can support decision-making in this context by early identification of donor organs at high risk of discard, for instance to enable timely interventions to improve organ utilization such as rescue allocation. Although various ML models have been applied, their results are difficult to compare due to heterogenous datasets and differences in feature engineering and evaluation strategies.

**Objective:** This study aims to provide a systematic and reproducible comparison of ML models for donor-level kidney discard prediction with an emphasis on methodological benchmarking rather than clinical implementation. Here, discard is defined as bilateral discard (both kidneys discarded) and transplantation indicates at least one kidney transplanted.

**Methods:** We trained five commonly used ML models: Logistic Regression, Decision Tree, Random Forest, Gradient Boosting, and Multilayer Perceptrons along with an ensemble model on data from 4,080 deceased donors (death determined by neurologic criteria) in Germany. We implemented an unified benchmarking framework, including standardized feature engineering, supervised feature selection, and Bayesian hyperparameter optimization. Model performance was assessed for classification and discrimination (AUC, MCC), calibration (Brier score, calibration slope and intercept), and explainability (SHAP).

**Results:** The ensemble achieved the highest classification and discrimination performance (AUC 0.87, 95% CI 0.834-0.896, and MCC 0.542, 95% CI 0.469-0.614). However, the results did not differ significantly from those of Logistic Regression. Individual models such as Logistic Regression, Random Forest, and Multilayer Perceptrons performed comparably and substantially better than single Decision Trees. Platt scaling improved calibration for tree-based and neural network-based models, while Logistic Regression was already well calibrated. SHAP consistently identified donor age and renal function markers as dominant predictors across models, reflecting clinical plausibility and interpretability.

**Conclusion:** This study demonstrates that consistent data preprocessing, feature selection, and evaluation can be more decisive for predictive success than the choice of the ML algorithm. The proposed benchmarking framework enables transparent comparison of models and provides a reproducible foundation for improving donor-level kidney discard prediction.

## Introduction

The global shortage of donor organs remains a significant challenge in modern medicine. While the demand for transplants continues to rise, organ donations fluctuate and often fall short of clinical needs [1]. At the same time, the quality of donor organs has declined, with increasing donor age and a reduced number of usable organs per donor [2]. These circumstances make allocation and acceptance decisions both highly complex and time-critical as preservation limits and ischemic-time related risks impose strict time constraints. As a consequence a substantial proportion of retrieved donor kidneys is ultimately discarded. Therefore, a research question of particular relevance in this context is the early identification of donor organs at high risk of being discarded, for instance to enable interventions to increase organ utilization such as timely rescue allocation efforts or applying machine perfusion to improve organ quality.

Machine learning (ML) models can support clinical decision-making by making use of the large volumes of available data, including biopsy reports, laboratory values, and free-text clinical records. Consequently, various ML models have been applied in the context of organ transplantation, often with promising results [3–9]. Thereby, kidney transplantation represents a frequently studied area due to the relatively large number of available cases. Commonly used ML models include Decision Trees (DT) and Regression Models as well as more flexible methods such as Random Forests (RF), Gradient Boosting, and artificial neural networks, which are generally less interpretable but may offer higher predictive performance. These models also have been applied to the specific task of predicting kidney discard risk, yielding heterogeneous results: Logistic regression (LR) remains one of the most commonly used approaches in this context, identifying key predictors associated with organ discard, such as biopsy findings and high Kidney Donor Profile Index scores [5,9–15]. Some studies report improved predictive performance when using more advanced ML models, such as gradient boosting or RF [7,14]. Other studies have found ML approaches to perform similarly to LR [15]. Very recently, Guan et al. [16] developed prediction models to assess whether current out-of-sequence placements of the U.S. allocation policies are appropriately focused on organs at risk of nonuse. However, these studies are difficult to compare directly, as the underlying datasets vary substantially. For example, some studies analyze a large national dataset with several hundred predictors, whereas another relied on single-center data with a much smaller sample size and a limited set of predictors. Furthermore, the role of feature selection [17] is often overlooked, potentially favoring models with built-in feature selection mechanisms such as RF. This makes it difficult to draw systematic conclusions about which models are most suitable for predicting kidney discard risk and thus supporting clinical decision-making in organ transplantation. The aim of this study was to address this gap by systematically comparing different ML approaches for predicting donor-level kidney discard risk (specifically identifying donors whose kidneys are both discarded versus those where at least one kidney is utilized), using a consistent preprocessing and transformation pipeline on a shared dataset of kidney transplantation candidates.

This work makes two main contributions. First, multiple state-of-the-art ML models are systematically evaluated with advanced techniques such as model calibration and Bayesian hyperparameter optimization applied to ensure an objective comparison. Second, not only prediction performance but also model explainability is compared through the use of SHAP, which enables the identification of key predictors and supports clinical interpretability.

Although the proprietary data used in this study cannot be shared because of confidentiality restrictions, we provide comprehensive access to all accompanying scripts for data processing, feature engineering and selection, model optimization, training, evaluation, calibration, and visualization to promote transparency and reproducibility.

The structure of this paper is as follows. Section 2 describes the data, the ML models, and feature engineering strategies, followed by an outline of the model evaluation methodology. Section 3 presents the results of the model comparison, including classification and discrimination performance, model calibration and explainability. Finally, Section 4 concludes the paper and highlights directions for future work.

# Methods

## Prediction Target and Study Objective

The prediction target was defined at the donor level. Each donor was classified according to the final utilization outcome of their kidneys: a donor was classified as transplanted if at least one kidney was transplanted and as discarded if neither kidney was transplanted. Although organ-level prediction may be more clinically relevant, many of the available attributes refer to the donor rather than a specific kidney. For this reason, the outcome was defined at the donor level in consistency with the available data. Hence, we predict the final outcome of the entire decision process leading to the acceptance or complete discard of a donor's offered kidneys. This final outcome may result from decisions made by multiple potential recipients and their healthcare providers who are sequentially allocated.

The study follows a comparative modelling approach, in which different methods are evaluated and benchmarked rather than a single model being developed. Identifying promising modelling approaches for predicting the probability that a kidney is accepted or discarded may support future work on developing a deployable predictive model as part of an early alert system for organizations involved in organ allocation, potentially enabling timely interventions such as rescue allocation for hard-to-place organs.

Figure 1 illustrates the overall methodological workflow. The dataset (n=4,080) was initially divided into a training set (80%, n=3,263) and an independent test set (20%, n=817). Preprocessing, including imputation and feature engineering, was performed using the training data, with the resulting parameters applied to the test set. Algorithm-specific feature selection was conducted on the training data using 500 trials with 10 inner hyperparameter optimization trials, and three-fold cross-validation. Subsequently, hyperparameters were optimized on the optimized feature sets using 100 trials and five-fold cross-validation.

For final model development, the training data were split into fitting (85%, n=2,773) and validation (15%, n=490) sets. The fitting set was used for model training and feature scaling, while the validation set supported early stopping, calibration, and threshold optimization. Final performance was assessed on the independent test set using classification and discrimination metrics, calibration measures, and SHAP-based explainability analyses.

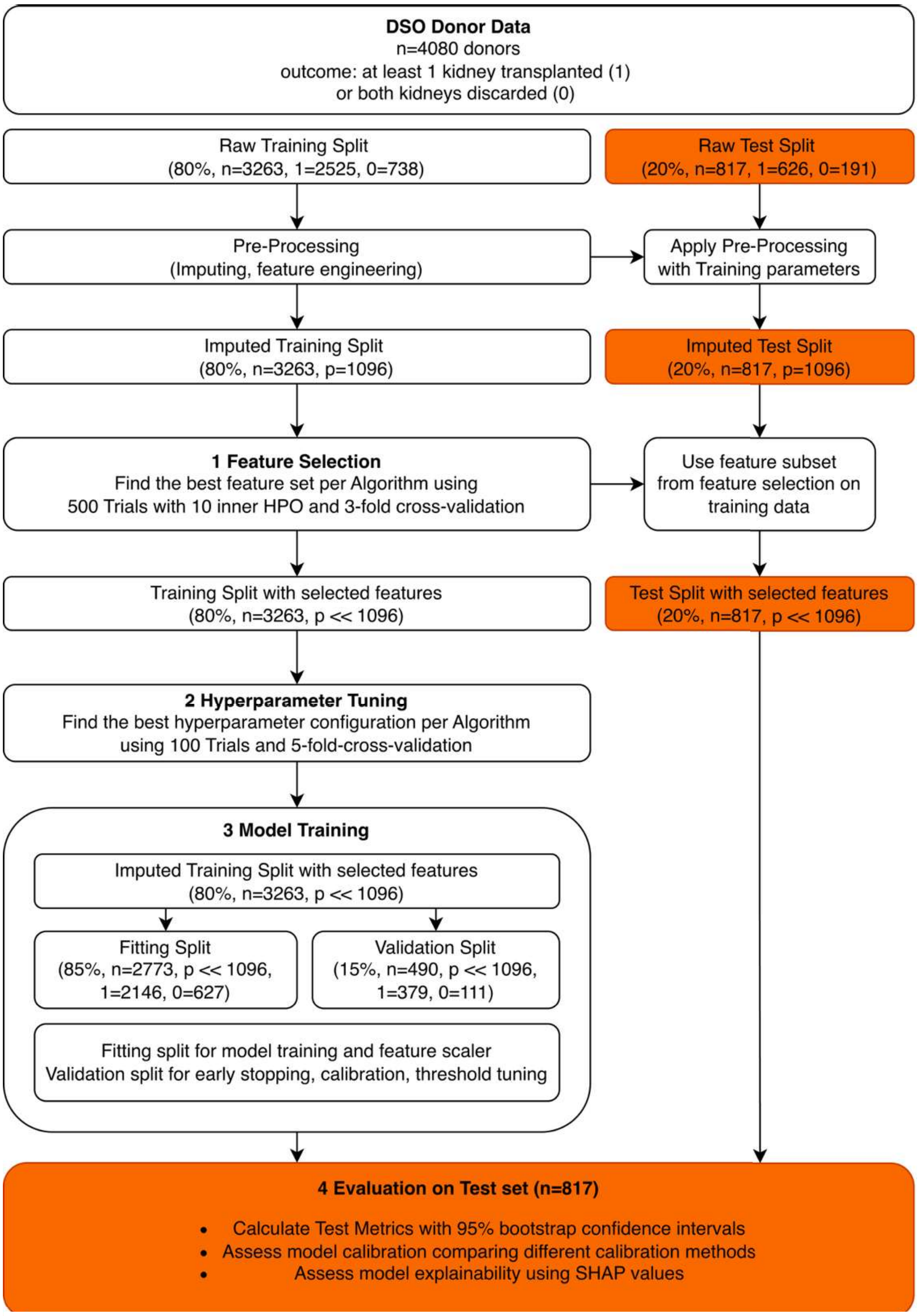

**Figure 1:** Study design. Donors were split into a training set (80%) and a held-out test set (20%). Preprocessing, feature selection, hyperparameter tuning, and model training used only the training data, and preprocessing parameters were applied unchanged to the test set. The test set (orange) was used exclusively for the final evaluation. n, number of donors; p, number of features.

Two guidelines are partially relevant but cannot be fully applied due to limited alignment with the research question, although they are taken into account where appropriate: The TRIPOD-AI [18] has been published to provide reporting recommendations for studies that develop or evaluate the performance of a prediction model. The TRIPOD-AI refers to prediction models for individual patient prognosis or diagnosis, while our prediction does not refer to a patient population but to organs offered for transplantation. Moreover, the aim of this study was not to derive a single prediction model for subsequent clinical deployment but to systematically compare different ML approaches for predicting kidney discard risk, using a consistent preprocessing and transformation pipeline. Nevertheless, we followed the TRIPOD-AI guideline where applicable, particularly with regard to the reporting of Methods and Open Science components. The 'predictimand framework' [19] is not directly applicable here, as we do not consider a treatment or time-dependent event. However, its recommendation to explicitly define the prediction target has been followed at the beginning of this section. The completed TRIPOD+AI checklist is provided in the supplement materials.

## Data

Data on all kidneys offered for transplantation from deceased adult donors in Germany between April 2020 and August 2024 were used. The dataset, provided by the German Organ Procurement Organization (DSO), includes information from $n = 4080$ adult donors (all deaths determined by neurological criteria). Only donor-related characteristics are considered in predicting whether a kidney would discarded, while recipient characteristics are intentionally excluded. This approach enables the identification of donor kidneys at high risk of discard based solely on information available at the time of the organ offer, prior to allocation to a specific recipient.

Donor data were provided by the German Organ Procurement Organization (DSO) as an independent research database for secondary analysis. All identifiers were removed, and only anonymized data are included. The dataset includes those data that the DSO is legally required to provide to the national transplantation registry under national transplantation law (TPG). The secondary analysis of these anonymized data was conducted in compliance with the transplantation law.

The outcome was defined at the donor level rather than the individual kidney level. A donor was classified as "discarded" if neither kidney was transplanted and as "transplanted" if at least one kidney was transplanted. This donor-level approach was adopted because many of the available attributes refer to the donor rather than to a specific kidney and also ensured independent observations and circumvent intra-individual correlations. In most cases (3,716 out of 4,080), both kidneys from a donor were either transplanted or discarded, making a donor-level classification straightforward. In 364 cases, only one kidney was transplanted while the other was discarded. These cases are classified as "transplanted" at the donor level. In summary, 929 (22.8%) of all donors are classified as discarded and 3151 (77.2%) of all donors are classified as transplanted (i.e. not discarded).

Donors providing multiple organs are not excluded as organ combinations do not interact with kidney discard. They may influence post-transplant function, which was however not the focus of this study.

The data comprises results from examinations of various body parts, information about any medication that had been prescribed, laboratory measurements from the domains of virology, pathology and microbiology as well as details on hospital stays and medical diagnoses including ICD-10 codes (International statistical classification of diseases and related health problems) [20]. In summary, p=1096 variables are available and were further processed in the feature engineering step (see section 3.3). A comprehensive description of the study cohort across all these variables is not feasible. Therefore, we extract the features that will later turn out to be most important for prediction and summarize them in Table 1.

Table 1: Descriptive statistics of the most influential features as reported in Figure 6. Individual dummy variables from one-hot encoding were excluded, as they do not capture the full information of the original categorical variable.

| Feature | All donors $N = 4080$ | Transplanted $N = 3151$ | Discarded $N = 929$ |
|---|---|---|---|
| Variables originally reported: | | | |
| Age (years), median [IQR] | 58.0 [46.0, 69.0] | 56.0 [44.0, 66.0] | 68.0 [56.0, 78.0] |
| Diuresis ( $ml/h/kg$(body weight)), median [IQR] | 1.54 [1.01, 2.22] | 1.61 [1.11, 2.38] | 1.27 [0.63, 1.79] |
| Variables extracted from time series: | | | |
| eGFR_MDRD ($ml/min/1.73m^2$), median [IQR] | | | |
| - minimum | 69.68 [44.95, 95.01] | 76.41 [53.84, 100.41] | 39.53 [19.81, 67.13] |
| Creatinine ($\mu\ mol/l$ ), median [IQR] | | | |
| - Minimum | 0.78 [0.6, 1.0] | 0.72 [0.58, 0.9] | 1.02 [0.77, 1.5] |
| - Maximum | 97.2 [74.22, 142.0] | 88.4 [70.7, 123.8] | 153.8 [97.2, 291.7] |
| - Intercept | 81.59 [61.93, 116.0] | 75.92 [59.47, 101.52] | 119.0 [84.0, 197.31] |
| - Last | 79.6 [61.9, 114.9] | 75.1 [59.2, 99.0] | 119.0 [79.6, 193.6] |
| - Standard deviation | 11.24 [5.77, 22.54] | 10.3 [5.36, 18.81] | 19.4 [7.45, 59.77] |
| Nitrite, median [IQR] | | | |
| - Count | 1.0 [1.0, 1.32] | 1.0 [1.0, 1.32] | 1.0 [1.0, 1.33] |
| Urea ($mmol/l$), median [IQR] | | | |
| - Minimum | 4.3 [3.0, 5.94] | 4.0 [2.8, 5.3] | 5.9 [4.2, 8.0] |
| - Maximum | 7.0 [5.0, 10.2] | 6.3 [4.7, 9.0] | 10.0 [6.8, 16.3] |
| Blood gases FIO2 (1/1), median [IQR] | | | |
| - First | 0.5 [0.35, 0.82] | 0.5 [0.35, 0.88] | 0.5 [0.4, 0.74] |
| Blood gases pH, median [IQR] | | | |
| - Count | 2.0 [1.0, 3.0] | 2.0 [1.0, 3.0] | 2.0 [1.0, 2.36] |
| Potassium ( $mmol/l$ ), median [IQR] | | | |
| - Maximum | 4.5 [4.2, 4.9] | 4.5 [4.2, 4.9] | 4.7 [4.4, 5.2] |
| INR (1/1), median [IQR] | | | |
| - First | 1.1 [1.0, 1.25] | 1.1 [1.0, 1.22] | 1.14 [1.05, 1.3] |
| Nitrite negative, n (%) | | | |
| - First | 3499 (85.8) | 2830 (89.8) | 669 (72.0) |
| Glucose negative, n (%) | | | |
| - First | 2928 (71.8) | 2392 (75.9) | 536 (57.7) |
| Protein qualitative negative, n (%) | | | |
| - First | 2069 (50.7) | 1772 (56.2) | 297 (32.0) |

The dataset was split based on donor id allocating 80% of donor ids to the training set and 20% to the test set to ensure independence between training and evaluation data. Feature Engineering was executed on both sets separately with thresholds calculated on only the training set and applied on the test set. Additionally, a fixed 10% subset of the training data was reserved as a dedicated validation set for final model training and calibration, ensuring strict separation from the held-out test set used for final evaluation.

## Feature Engineering

In this section, we describe feature transformations including time series processing, redundancy reduction, encoding of categorical variables and imputation of missing values. Some additional transformations of specific features are described in Supplement B.

Variables were selected, extracted and engineered without prior filtering by medical advisors to avoid selection bias. Only in cases of uncertainty primary investigators consulted medical experts to ensure appropriate data standardization (e.g. donor urine output adjusted to body weight of donor and standard time interval as $ml/kg/h$). Secondary validation was performed by medical advisors to ensure that the data reflect real world scenarios in deceased organ donation procedures.

### Feature Extraction from Time Series

Many of the laboratory measurements come in the form of time series. 49 variables were identified as time series. An additional 40 variables were not considered time series, as more than 50% of donors had only a single recorded value for these variables. For these we chose either the most recent or the first recorded value. We categorized the time series variables into the following subtypes to maximize the extracted information for each variable:

- Type 1: Time series variables with categorical values, that indicate a positive or negative outcome only.
- Type 2: Time series variables with numerical values and a mean number of observations per donor lower than two.
- Type 3: Time series variables with numerical values and a mean number of observations per donor of two or more.

For each time series variable, we extracted the following features for every donor: the first recorded value, the last recorded value, the number of entries, and the time span between the first and last entry (measured in hours).

For time series variables with numerical values (type 2 and 3) we also extracted the standard deviation, minimum, and maximum values.

For time series variables of type 3, we further extracted features describing the linear trend over time: A linear regression per donor was performed modeling the value of the variable (dependent variable) over time measured in hours since the first entry (independent variable). Thus, the linear regression model for every donor $d$ is:

$$y_{d,i} = \beta_{d,0} + \beta_{d,1} \cdot t_i + \varepsilon_{d,i}$$

where $y_{d,i}$ denotes the value of the variable at the time point $t_i$. The estimated coefficients $\hat{\beta}_{d,0}$ and $\hat{\beta}_{d,1}$ were used as additional features.

Previous studies use the last recorded value or the minimum value [21,22]. Our additionally extracted features may capture changes over time: The number of entries may indicate that variable values for a donor have changed over time. Similarly, the time span between the first and last entry can also reflect temporal dynamics in the donor's condition. The standard deviation also provides insights into the level of variability. The estimated linear regression coefficients include information on whether the values generally increased or decreased over time and to what extent. In summary, a total of 364 features were extracted from the time series data.

**Feature Extraction from Donor Medication Data**

In the raw data there are 1772 unique medication names. These are structured such that the first word describes the medication and the remaining ones give additional information such as dosage. Considering only the first word of each medication entry reduced the number of medication names to 1103. Afterwards, the 40 most frequent medication names were selected for the final feature set. This selection captures approximately 80% of all medication administrations in the dataset. They were turned into binary features indicating, whether a donor had taken the medication or not.

**Categorical Variable Encoding**

We used One-Hot encoding for categorical variables with more than two values. Thereby, for variables containing ICD-10 codes, we replaced codes with a relative frequency of less than 1% of donors in the training set by missing values.

**Missing Value Imputation**

Approximately 14% of the values in our data are missing with 628 of the variables having incomplete data. These were imputed in the same way in training and test set unless stated otherwise as follows:

- 250 variables could be imputed using logical dependencies following expert advice. These include variables on certain diagnoses where a missing value can be interpreted as a negative value (i.e. no diagnosis present).
- For some categorical variables missing values were kept and considered a separate category. These are variables where a missing value is considered to potentially influence the decision about discard (e.g. 'urine_glucose').
- For two variables ('CPR duration' and 'ecmo') free text variables provided additional information and were used for imputation.
- 30 continuous variables describing physiological parameters were imputed with random samples from a normal distribution with mean and standard deviation of each variable, retaining only values from within the central 95% range of this distribution. The parameters of the normal distribution were estimated on the training set only and used to impute missing values in both the training and the test set.

- For 20 variables with a high percentage of missing values (>70%) in the training set the specific values were dropped and the variables dichotomized into "missing" and "not missing". In these cases imputation based on sparse data was considered unreliable and therefore not performed.
- The remaining 328 variables with missing values in the training set were imputed with an iterative imputing algorithm [23] using ridge regression to predict the missing values for both training and test set. The iterative imputer was fitted on the training set only and subsequently used to impute missing values in both the training and the test set. If there were variables with missing values in the test set that had no missing values in the training set, these were filled with the mean of that variable. This method was chosen to account for complex relationships between variables.

## Modeling Approach

The following section outlines our modeling approach, including the ML algorithms, feature selection techniques, and hyperparameter optimization methods applied to the kidney transplantation dataset.

### Machine Learning Algorithms

The following ML algorithms are trained to predict the probabilities of donor organs being either discarded or transplanted.

#### Decision Trees

Decision Trees (DTs) [24] are supervised learning models used for classification and regression tasks. They split data recursively into subsets based on feature thresholds, forming a tree-like structure where each internal node represents a decision and each leaf node provides a final prediction. DTs are known for their interpretability, as each split is based on simple rules derived from an impurity measure such as the Gini impurity, but are prone to overfitting, especially when the tree is deep.

#### Logistic Regression

Logistic Regression (LR) [25] is a ML algorithm from the linear family, commonly used for binary classification tasks. LR models the probability of an outcome by applying a sigmoid (logistic) function to a linear combination of input features. The model parameters are estimated by maximum likelihood, i.e., by minimizing the log-loss (binary cross-entropy) between predicted probabilities and observed outcomes, typically using an iterative optimization procedure. An additional advantage is its interpretability, since feature importance can be directly inferred from the model coefficients.

### Random Forest

Random Forest (RF) [26] is an ensemble learning method that builds multiple DTs and combines their outputs to improve predictive accuracy and decrease overfitting. The forest model uses the bagging method by training each tree on a random subset of the data and features resulting in reduced overfitting compared to individual DTs. An additional advantage is the possibility to derive aggregated feature importance measures across all trees, which enhances interpretability.

### eXtreme Gradient Boosting

Extreme Gradient Boosting (XGB) [27] is an ensemble learning algorithm based on gradient boosting [28], suitable for classification and regression tasks. The model builds DTs sequentially, where each new tree corrects the errors of the previous ones, optimizing performance through gradient descent. XGB performs especially well on structured data [29] and can even outperform deep learning algorithms [30]. An additional advantage is the availability of feature importance measures which provide interpretability despite the model's complexity.

### Multilayer Perceptron

Multilayer Perceptron (MLP) (also known as Deep Learning) [31] is an artificial neural network used for both classification and regression tasks. It consists of multiple layers of interconnected neurons, including an input layer, one or more hidden layers and an output layer, where each neuron applies a weighted sum of inputs followed by a nonlinear activation function. MLPs are capable of capturing complex, nonlinear relationships but require careful tuning of hyperparameters and large datasets to generalize well. Although they are more commonly applied to unstructured data such as images or text, MLPs can also be used on structured tabular data.

### Ensemble

An ensemble model, referred to as the data ensemble model (DE), was implemented to combine the predictions of several pretrained classifiers. In this study the DE integrates the outputs of LR, XGB, RF and MLP. Each model produces a probability for both classes and the ensemble aggregates these outputs by computing the mean of the predicted probabilities across models. The final classification result is determined by selecting the class with the highest averaged probability. DTs were excluded from the ensemble due to their comparatively lower performance, as will be demonstrated in later sections.

### Feature Selection

To address the high dimensionality of our dataset, we apply a feature selection approach to identify the most promising subset of features. This reduces redundancy, improves model interpretability, and prevents overfitting by focusing on the variables most relevant to prediction [17]. Because different machine learning models capture relationships in different ways, we perform the feature selection separately for each model but always with the same method. This ensures that each algorithm operates on a feature set tailored to its specific inductive biases and complexity. To capture nonlinear and multivariate interactions between features, we employ an optimization method based on genetic algorithms [32,33]. This approach can uncover complex relationships that simpler univariate filters (e.g., correlation or chi-square tests) may miss [34,35]. Feature selection is performed exclusively on the training set, so the test set remains untouched until the final evaluation.

The general approach is applied separately for each ML model as depicted in Figure 2 and described in the following.

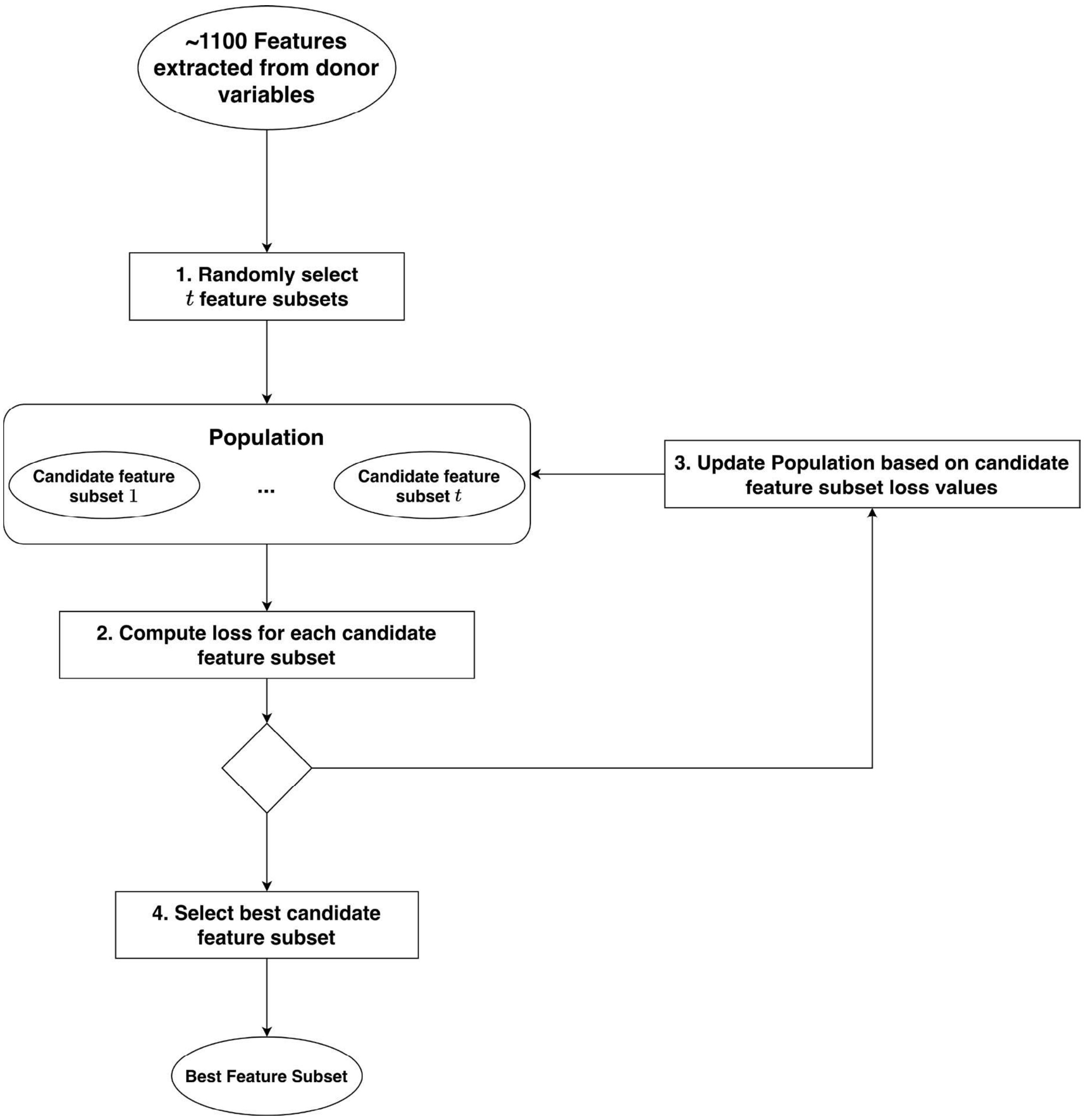

Figure 2: Schema of the feature selection process. Starting from the full feature space, multiple candidate feature subsets are created and evaluated by training and validating a supervised model. Each subset is assigned a loss value based on its predictive performance, and the subset with the smallest loss is selected as the final feature subset.

We use the Nondominated Sorting Genetic Algorithm II (NSGA-II) [36] for single-objective feature selection optimization, as it natively handles binary and categorical parameters, unlike other evolutionary strategies [37]. We use the optuna [38] implementation for the NSGA-II algorithm.

1. First, a population of 50 randomly selected feature subsets are used as candidates. Feature sets here are represented as boolean vector of length ~ 1100 (number of features in the feature set) where each element represents a feature and is set to either one or zero depending on whether the feature is included in the subset or not.
2. Next, for each feature subset, we train the ML algorithm using only the corresponding features. Performance is evaluated with a loss metric:
    - We perform a nested hyperparameter optimization to build the model architecture: For each candidate feature subset we run a randomized hyperparameter search [39] with 10 trials to avoid overfitting of the default model hyperparameters to a specific feature subset. For each of these 10 trials, we evaluate the generated model configuration with 3-fold cross-validation on the training set.
    - The yielded validation loss from the 3-fold cross-validation is calculated using the Matthews Correlation Coefficient (MCC) [40], a balanced metric that combines true/false positives/negatives into a single correlation score providing a more informative assessment than other common metrics for imbalanced settings [41,42].
    - The overall loss for a subset is defined as the negative mean 3-fold cross-validation MCC of the best of the 10 hyperparameter configurations from the inner search, plus an explicit penalty term on the feature count, $\lambda \cdot n_{\text{features}}$ (with $\lambda = 0.0005$), to discourage the selection of large, redundant feature subsets.
3. Finally, the best performing feature subsets are selected and form the basis for evolutionary operations (selection, crossover, and mutation) which yield a next generation of the population of feature subsets. For these, the process is repeated.
4. After 5000 trials, i.e. 500 evaluated candidate feature subsets (20 generations of 50 candidates each, amounting to approximately 500 × 10 × 3 = 15000 model fits per ML algorithm), the process terminates, yielding the best-performing feature subset for each ML algorithm.

Our approach follows prior work in related domains but differs in the use of evolutionary search per model and the nested randomized-architecture evaluation strategy, which together aim to produce robust, compact feature sets tailored to each classifier rather than a single global subset [43,44].

Additionally, 93 features that were redundant or constant after imputation were dropped.

### Model Hyperparameter Optimization

The feature selection produces five optimized feature subsets, one tailored to each ML model. To identify the best architecture for each algorithm, we perform a dedicated hyperparameter optimization on the corresponding optimized feature space.

We use optuna with the Tree-structured Parzen Estimator (TPE) [45] sampler as a Bayesian hyperparameter strategy [46] to identify the best hyperparameter configurations for each ML algorithm. The optimization objective was the mean MCC across a 5-fold cross-validation performed exclusively on the training set. The test set was not used at any stage of the optimization. For each algorithm we run 100 trials. The corresponding hyperparameter search spaces are listed in Supplement Tables D1-D5 and the selected (best) architectures are reported in Supplement Tables D6-D10. Supplement Figure D1 shows the outer cross-validation performance per trial.

## Model Evaluation

The following sections present the concrete evaluation steps of our model assessment. We first report how we measure classification and discrimination performance, then present methods for calibration analyses on predicted probabilities, and finally summarize how we get explainability findings using SHAP-based attributions.

In line with the STRengthening Analytical Thinking for Observational Studies (STRATOS) initiative [47], our primary measures focus purely on statistical performance. Metrics that conflate statistical and decision-analytic evaluation are not used for model selection or tuning. Accordingly, the primary performance measure is the area under the receiver operating characteristic curve (AUC) [48], a semi-proper measure of discrimination that is part of the STRATOS core set. Although the AUROC has been criticized as overly optimistic under class imbalance [49,50], STRATOS argues that this criticism conflates class imbalance with misclassification costs. For feature selection, hyperparameter tuning and threshold selection, we use the MCC (Equation 1) [40], a stringent measure that incorporates all four cells of the confusion matrix and yields high values only if both classes are predicted well.

We additionally report the area under the precision–recall curve (PRAUC) for both outcome classes, each treated in turn as the positive class. We report the PRAUC for descriptive purposes only, as STRATOS advises against its use since it ignores true negatives and depends on outcome prevalence. Sensitivity and specificity, as well as positive and negative predictive values (PPV and NPV), are reported as secondary measures in Supplement E2 and, in line with STRATOS, always in pairs.

We do not report clinical utility measures such as net benefit, although they are part of the STRATOS core set, because our objective is statistical model evaluation and comparison rather than an assessment of clinical utility. Our results should therefore not be interpreted as evidence of clinical benefit.

Equation 1

$$MCC = \frac{TP \cdot TN - FP \cdot FN}{\sqrt{(TP + FP)(TP + FN)(TN + FP)(TN + FN)}}$$

**Model Training and Estimation of Test Variance**

Each algorithm is trained once with its selected feature subset, its selected hyperparameter configuration and a fixed random seed. The models are fitted on 85% of the training set. The remaining 15% are used as validation data for early stopping (XGB, MLP), calibration and threshold selection (See Table 2). Z-score scaling parameters (mean and standard deviation of the respective feature) are computed exclusively on the training set and applied to both, the validation and test sets, to prevent information leakage. The held-out test set is only used for the final evaluation.

Donors with at least one transplanted kidney are classified as 1 ("positive"), whereas donors whose kidneys were both discarded are classified as 0 ("negative"). A true positive therefore corresponds to a correct prediction that at least one of the donor's kidneys was transplanted, and a true negative to a correct prediction that both kidneys were discarded. For classification metrics, the predicted class depends on a threshold of the predicted probability of transplantation. For each model, we derive a threshold on the validation split at which the proportion of donors predicted as transplanted equals the observed prevalence of transplantation. This threshold is used only if it yields a higher validation MCC than the default threshold of 0.5. The selected thresholds are reported in supplement Table E1.

All test-set measures are reported with 95% Bias-corrected and accelerated (BCa) bootstrap confidence intervals [51] based on 2,000 resamples of donors. All models are evaluated on the same resamples, so differences to LR, the reference model, are estimated as paired differences. We use them to identify significant differences compared to LR; a difference indicates a significant differences if the confidence interval does not include zero. However, results are to be interpreted in an exploratory manner as no correction for multiple testing is applied.

Table 2: Distribution of samples across training, validation, and test sets, by outcome.

| **Dataset** | **Samples** | **>= 1 kidney Transplanted=1** | **Both kidneys Discarded=0** |
|---|---|---|---|
| Training | 2773 | 2146 | 627 |
| Validation | 490 | 379 | 111 |
| Test | 817 | 626 | 191 |

## Model Calibration

The measures above evaluate discrimination and classification but do not show whether the predicted probabilities are accurate. The AUC depends only on the ranking of the predictions and the MCC only on whether they exceed a threshold, so a model can rank donors well and still systematically over- or underestimate the probability of transplantation [52]. We evaluate predicted probabilities using the balanced Brier score [53] and quantify calibration using the calibration intercept and slope [52]. The balanced Brier score is based on the Brier score [54], defined as the mean squared error between predicted probabilities and true binary outcomes (range $0 - 1$ where lower is better), but computes this error separately for both outcome classes and averages the two respective metrics (lower is better). Both outcome classes thus contribute equally despite the class imbalance. Consequently, the balanced Brier score favors probabilities shifted towards 50% and can increase after post-hoc calibration [55], which is why calibration is judged by the intercept, slope and reliability plots.

Post-hoc calibration adjusts a trained model's predicted probabilities to better match the true outcome frequencies, improving the reliability of its probability estimates.

Specifically, post-calibration methods transform the raw prediction scores such as the raw prediction scores of a classifier into well-calibrated probability estimates. We compare three standard post-calibration approaches. Platt scaling [56] fits a parametric sigmoid function (a two-parameter logistic model, commonly implemented as LR) that maps these scores to probabilities. Its parameters are optimized on the validation data and then applied to the test predictions. Isotonic regression [57], in contrast, provides a non-parametric monotonic mapping capable of modeling more complex distortions between raw scores and true probabilities. Temperature scaling [58] divides the logit of the predicted probability by a single fitted parameter. It corrects over- and underconfidence but has no intercept term to correct a systematic over- or underestimation of risk. In our setup, each calibrator is trained separately for each model, including the ensemble model (DE), on the validation predictions. It is then applied to the test set and evaluated using the balanced Brier score with 95% BCa bootstrap confidence intervals, the calibration intercept and slope, and reliability plots to visualize residual miscalibration. Calibration curves, intercept and slope refer to the predicted probability that at least one kidney is transplanted. We specified Platt scaling a priori as primary method because isotonic regression is prone to overfitting small calibration sets and temperature scaling lacks an intercept term.

Each algorithm is trained once with a fixed random seed, all calibration results refer to this final model.

## Explainability

To evaluate explainability we use SHapley Additive exPlanations (SHAP) [59] as a model-agnostic method that applies uniformly across model families. SHAP is grounded in Shapley values [60] from cooperative game theory. For each observation, it decomposes the difference between the model's predicted probability and the mean prediction over the background data into additive feature contributions. This additive property makes SHAP suitable for both, local (per-instance) and aggregated global interpretation. SHAP has been widely applied in tabular and clinical ML studies to highlight predictive drivers, support clinician interpretation and to improve transparency in medical prediction tasks [61,62].

We use the Permutation explainer for all models. The training observations serve as background data, and the explained output is the model's predicted probability of transplantation. SHAP values are computed for alle test observations. For model comparisons, we focus on global interpretation and summarize each feature by its mean absolute SHAP value. We report the 10 most important features per model with 95% BCa bootstrap confidence intervals of the mean absolute SHAP value (2,000 resamples of the test set).

For the ensemble DE, which averages the predicted probabilities of its members, we derive SHAP values from the additivity axiom of Shapley values [60]. Because all members are explained with the same background data, the SHAP value of a feature in DE equals the mean of its SHAP values in the members.

Model-specific explainers (eg, impurity-based feature importances of tree models or LR coefficients) can be suitable when a single model is interpreted, but they are not comparable across model families [63]. Impurity-based importances are, in addition, derived from the splits on the training data and favor features with many possible split points [63].We therefore report them only as a complement to SHAP.

**Ethical considerations**

The data were provided by the German Organ Procurement Organization (DSO) as an independent research database for secondary analysis. All identifiers were removed, and only anonymized data are included. The analysis of these anonymized patient data is permitted under §14 and §15 of the Transplantation Act (Transplantationsgesetz, BGBl. I S. 2206). Informed consent was not required according to §14 of the Transplantation Act (Transplantationsgesetz, BGBl. I S. 2206) as only retrospective and anonymized data from deceased organ donors were used. Medical ethical approval was not required according to §15 of the Professional Code of Conduct of the State Medical Association of Hesse as only retrospective and anonymized data of deceased organ donors were used. The research was conducted in accordance with the Declaration of Helsinki. No compensation was provided, as only anonymized data of deceased donors were analyzed.

## Results

In this section, we first present results of the feature selection and model performances. Then we analyze probability calibration before concluding with an assessment of model explainability.

### Feature Selection

Figure 3 summarizes the feature selection. Each search ran 500 trials in 10 generations (8 for LR). For all models, the mean cross-validated MCC per generation increased from the first to the last generation. The selected subsets contained 468 to 534 features depending on the model (See Supplement Figure C1 for intersections between models and Supplement Table C1 for intersecting features between all models). The cross-validated MCC exceeded that of the full feature set for each model. This suggests that the full feature set contains uninformative features and that a guided search such as NSGA-II helps to exclude them.

The best subsets agreed moderately. The 50 best trials of each algorithm shared more features than 50 random trials of the same search (mean Jaccard index 0.41–0.45 vs 0.35–0.37, random feature sets 0.29–0.32). A few features were used by almost all of the best trials and a few by almost none, but feature subsets with similar MCC generally differed substantially.

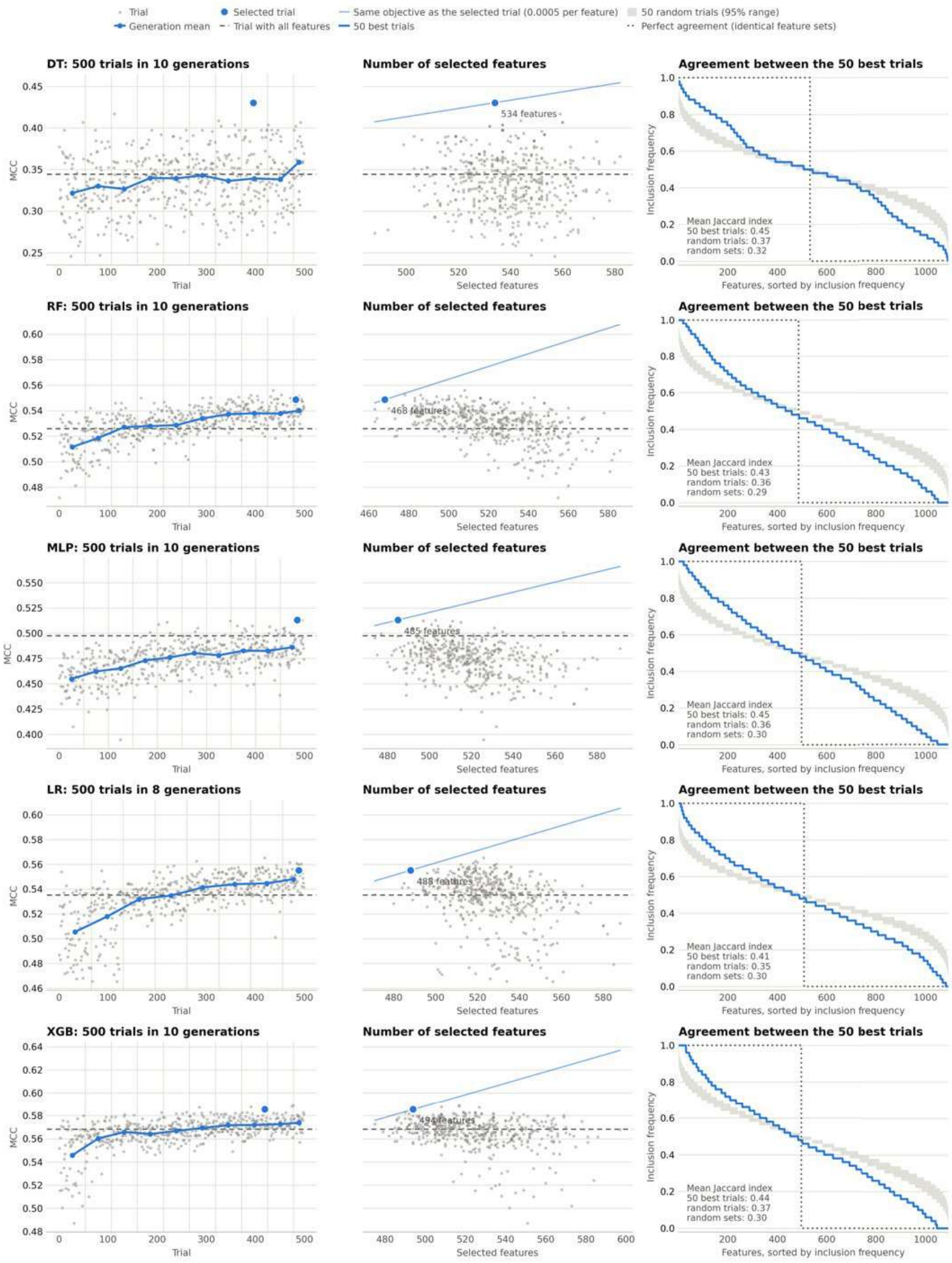

**Figure 3:** Evolutionary feature selection (NSGA-II) for each model, with MCC values from 3-fold cross-validation on the training set. **Left:** MCC of each trial, mean MCC per generation (vertical lines separate generations), the selected trial, and the trial with all features (dashed line). **Middle:** MCC versus number of selected features. The objective subtracts 0.0005 from the MCC for each selected feature. The light blue line indicates the required MCC for a given number of selected features to maintain the same objective value as the selected trial when accounting for the penalty in terms of number of features

**Right:** for each feature, the proportion of the 50 best trials that selected it (inclusion frequency), sorted from highest to lowest. The steeper the curve, the more the trials agree on their features. The gray band shows the 95% range of this curve for 50 trials drawn at random from the same run, and the dotted line shows perfect agreement, i.e., identical feature sets.

## Classification and Discrimination Performance

Figure 4 shows the test-set performance and Figure 5 shows the paired differences to LR. The ensemble DE reached the highest AUC (0.870, 95% CI 0.834 to 0.896), followed by LR (0.861, 95% CI 0.825 to 0.890), MLP (0.853, 95% CI 0.813 to 0.881), RF (0.849, 95% CI 0.814 to 0.877) and XGB (0.842, 95% CI 0.805 to 0.872). The MCC of these five models ranged from 0.490 (RF) to 0.542 (DE). For AUC and MCC, the confidence intervals of all their paired differences to LR included 0, although only narrowly for XGB in AUC (−0.019, 95% CI −0.037 to 0.000). The largest MCC difference was 0.032 for DE (95% CI −0.024 to 0.083). DT performed worse than LR on all measures, with an AUC of 0.684 (difference −0.177, 95% CI −0.226 to −0.135) and an MCC of 0.220 (difference −0.290, 95% CI −0.379 to −0.196).

The descriptively reported PRAUC showed a similar pattern (See supplement Figure E4 for the PRAUC curves). With transplantation as the positive class, it ranged from 0.928 (XGB) to 0.950 (DE) for the five models, compared with a prevalence of 0.766. With discard as the positive class, it ranged from 0.671 (XGB) to 0.728 (DE), about three times the discard prevalence of 0.234. For discards, RF (−0.040, 95% CI −0.077 to −0.007) and XGB (−0.050, 95% CI −0.090 to −0.014) had a lower PRAUC than LR, and XGB also had a narrowly lower one for transplantation (−0.016, 95% CI −0.036 to −0.000). DT again ranked last (0.852 and 0.425). Overall, no model was statistically significantly superior compared to LR in terms of AUC, MCC or PR-AUC.

The balanced Brier score ranged from 0.168 (DE) to 0.187 (LR) for the five models, compared with 0.25 for a constant prediction of 0.5. DE had a lower balanced Brier score than LR (−0.019, 95% CI −0.028 to −0.009), and XGB had a narrowly lower one (−0.014, 95% CI −0.027 to −0.000). DT had the highest balanced Brier score (0.285), worse than a constant prediction of 0.5. Confusion matrices, Sensitivity, specificity, PPV, and NPV are reported in supplement Figures E1-E3.

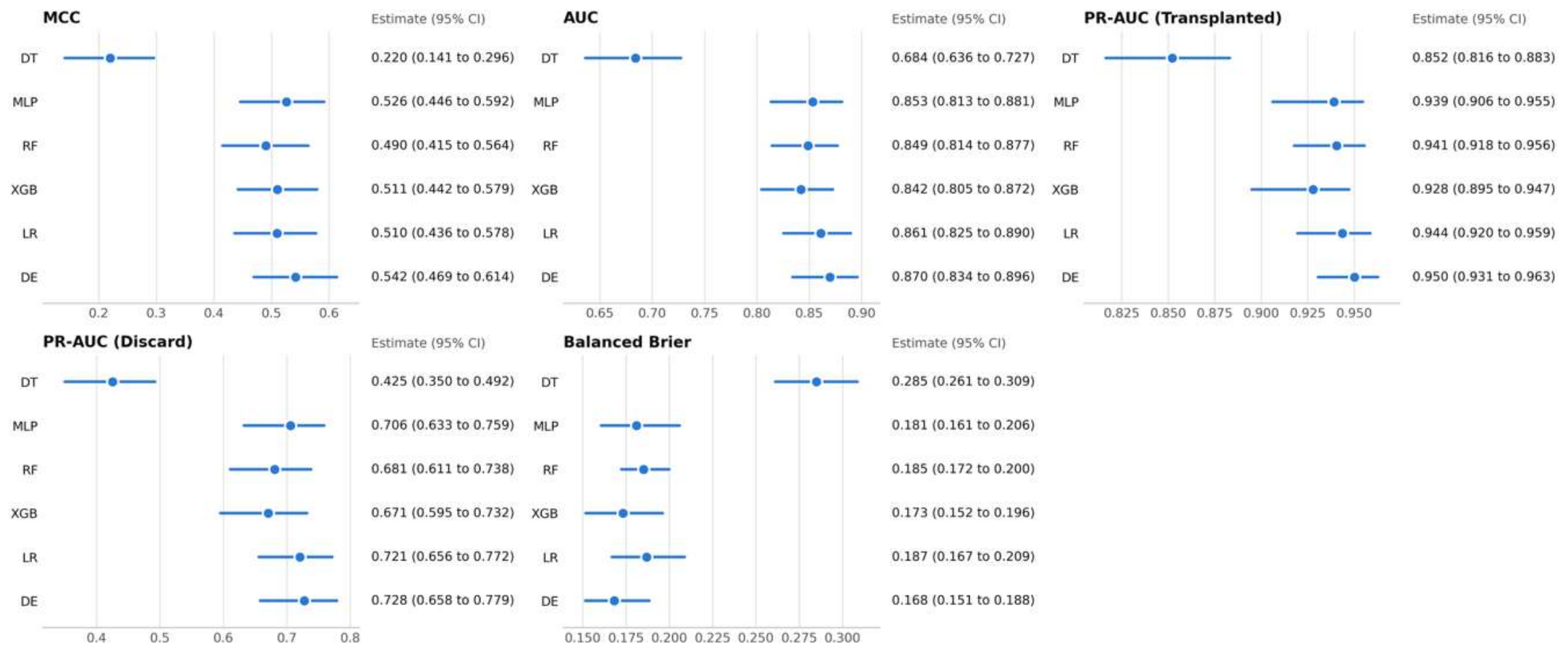

**Figure 4.** Test-set performance of all models (n = 817) with 95% BCa bootstrap confidence intervals (2,000 resamples). The MCC is computed at the thresholds selected on the validation split. PRAUC is shown with transplantation and with discard as the positive class. Lower balanced Brier scores are better.

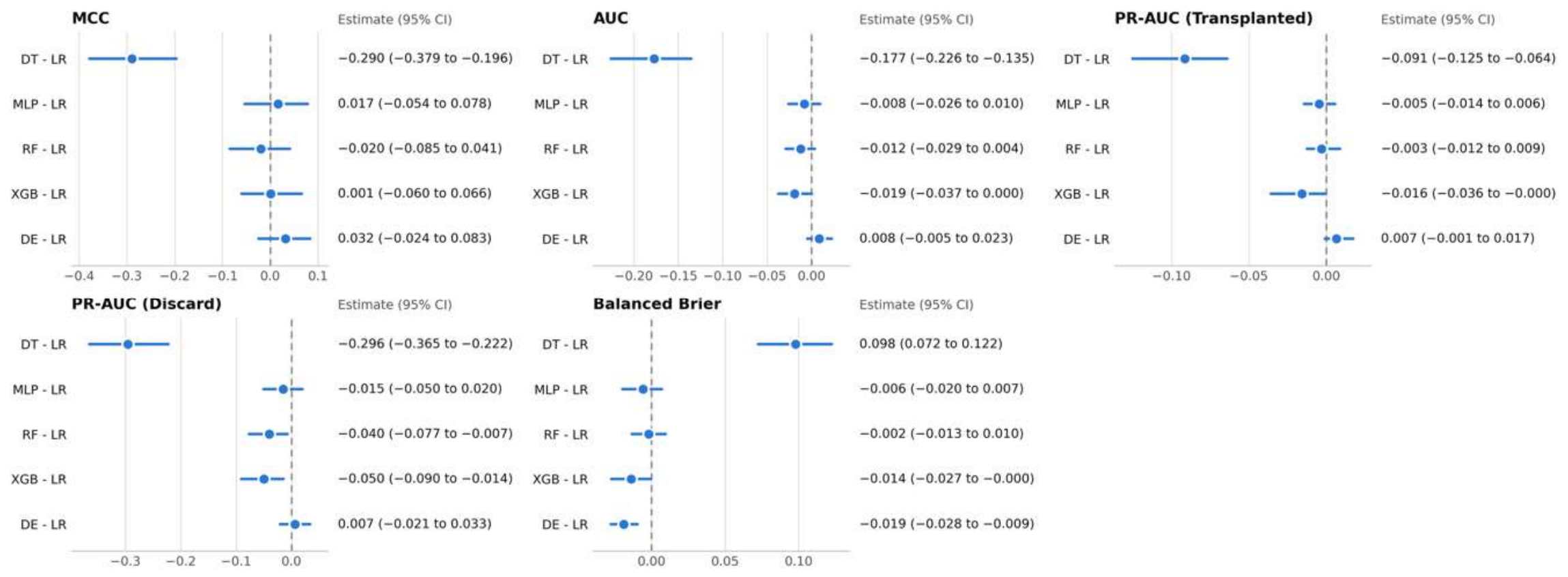

**Figure 5.** Paired differences to LR on the test set (model minus LR) with 95% BCa bootstrap confidence intervals (2,000 resamples). Intervals that exclude the dashed zero line indicate a difference. For the balanced Brier score, negative values favor the model.

Supplement Figure E5 compares the feature selection MCC with the cross-validated MCC from hyperparameter tuning and with the MCC on the validation split and the test set. Except for MLP, the test MCC at a threshold of 0.5 was lower than the cross-validated MCC of the selected subset, by 0.024 (LR) to 0.069 (XGB) and by 0.210 for DT. The MCC on the small validation split had wide confidence intervals and was higher than on the test set for all models.

Excluding the 70 test donors with one transplanted and one discarded kidney lowered the prevalence of transplantation from 76.6% to 74.4% and increased the AUC by at most 0.015, with the same ordering of the models by AUC. The calibration intercepts decreased with the lower prevalence, while the slopes changed little (Supplement Table E2).

**Model Calibration**

Figure 6 shows the calibration curves on the test set before and after Platt scaling, the primary calibration method. Without post-hoc calibration, only LR was well calibrated (intercept −0.01, slope 1.02), and MLP was mildly miscalibrated (intercept 0.19, slope 0.83). DT was strongly overconfident (slope 0.13), whereas RF (slope 1.82) and DE (slope 1.26) were underconfident. RF, XGB and DE also underestimated the probability of transplantation (intercepts 0.48, 0.54 and 0.32).

Platt scaling moved the slopes of all models to between 0.79 and 1.00 and their intercepts to between −0.28 and −0.02. The largest improvements occurred for DT (slope 0.13 to 0.98), RF (slope 1.82 to 0.88, intercept 0.48 to −0.28) and XGB (intercept 0.54 to −0.17), whose slope remained at 0.79. For LR, the slope remained close to 1 (1.02 to 1.00), but the intercept shifted from −0.01 to −0.14. All intercepts were negative after calibration, meaning that the models slightly overestimated the probability of transplantation.

The balanced Brier score increased after Platt scaling for all models, by 0.005 (DT) to 0.032 (XGB). As described in the Methods, this score favors probabilities shifted toward 50% and therefore does not reward calibration to the observed outcome frequencies. The Brier score, in contrast, decreased for DT (0.174 to 0.163), RF (0.133 to 0.122) and XGB (0.129 to 0.124) and changed by at most 0.001 for LR, MLP and DE.

Isotonic regression produced intercepts similar to those of Platt scaling but lower slopes (0.51 to 0.83), consistent with overfitting on the small validation split, and its Brier score was never lower than that of Platt scaling. Temperature scaling, which lacks an intercept term, gave inconsistent results. It brought the intercepts of RF (−0.07) and DE (0.01) close to 0, left those of XGB (0.49) and MLP (0.22) almost unchanged and worsened the calibration of DT (intercept 0.87, slope 0.47). Calibration curves and scores with confidence intervals for all investigated post-calibration methods are given in Supplement Figures F1-F5.

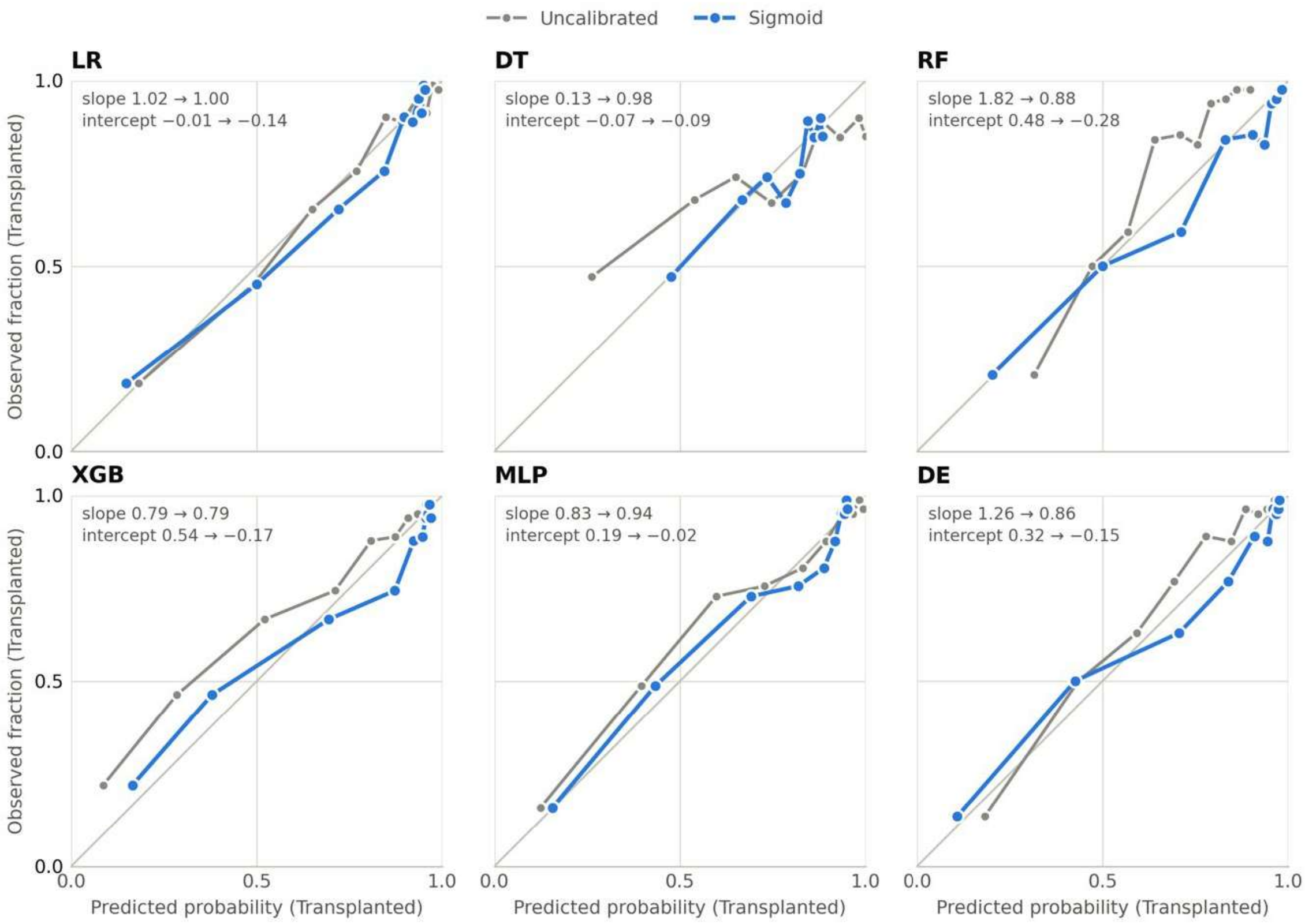

**Figure 6.** Calibration curves of all models on the test set without post-hoc calibration (grey) and after Platt scaling (blue) fitted on the validation split. Each point represents one of 10 bins with equal numbers of donors. The diagonal marks perfect calibration. Annotations give the calibration slope and intercept before and after calibration (ideal values 1 and 0).

## Explainability

Figure 7 shows the 10 features with the largest mean absolute SHAP value across all test samples for each model. Age was the most important feature in all models except DT, where it ranked second. In XGB, LR and DE, its mean absolute SHAP value was about twice that of the next feature, and its confidence interval did not overlap with that of any other feature. In MLP and RF, importance was distributed more evenly, and the confidence intervals of the leading features overlapped. In all models except DT, the second-ranked feature was a marker of renal function (creatinine or eGFR), and renal markers accounted for 3 (XGB) to 7 (MLP, DE) of the remaining nine features. These markers entered in different aggregations of their measurements, such as minimum, maximum, last value and intercept. Qualitative urine protein, urine output and an unremarkable kidney morphology on abdominal imaging also appeared among the top features of several models. DT deviated from this pattern. Its most important feature was a negative first urine hemoglobin test, and its top 10 contained echocardiographic, abdominal imaging and blood gas features but only one renal marker (maximum urea).

The model-specific importances (supplement Figures G1-G4) largely agreed with SHAP. Of their 10 most important features, 9 coincided with the SHAP top 10 for RF and LR, 7 for DT and 6 for XGB, and age ranked first or second in all four models. The rankings within the top 10 differed between the two approaches. For XGB, the model-specific importance additionally included two medication variables (metoprolol and regular medication intake) that did not appear among its SHAP top 10. No model-specific importance is available for MLP and DE.

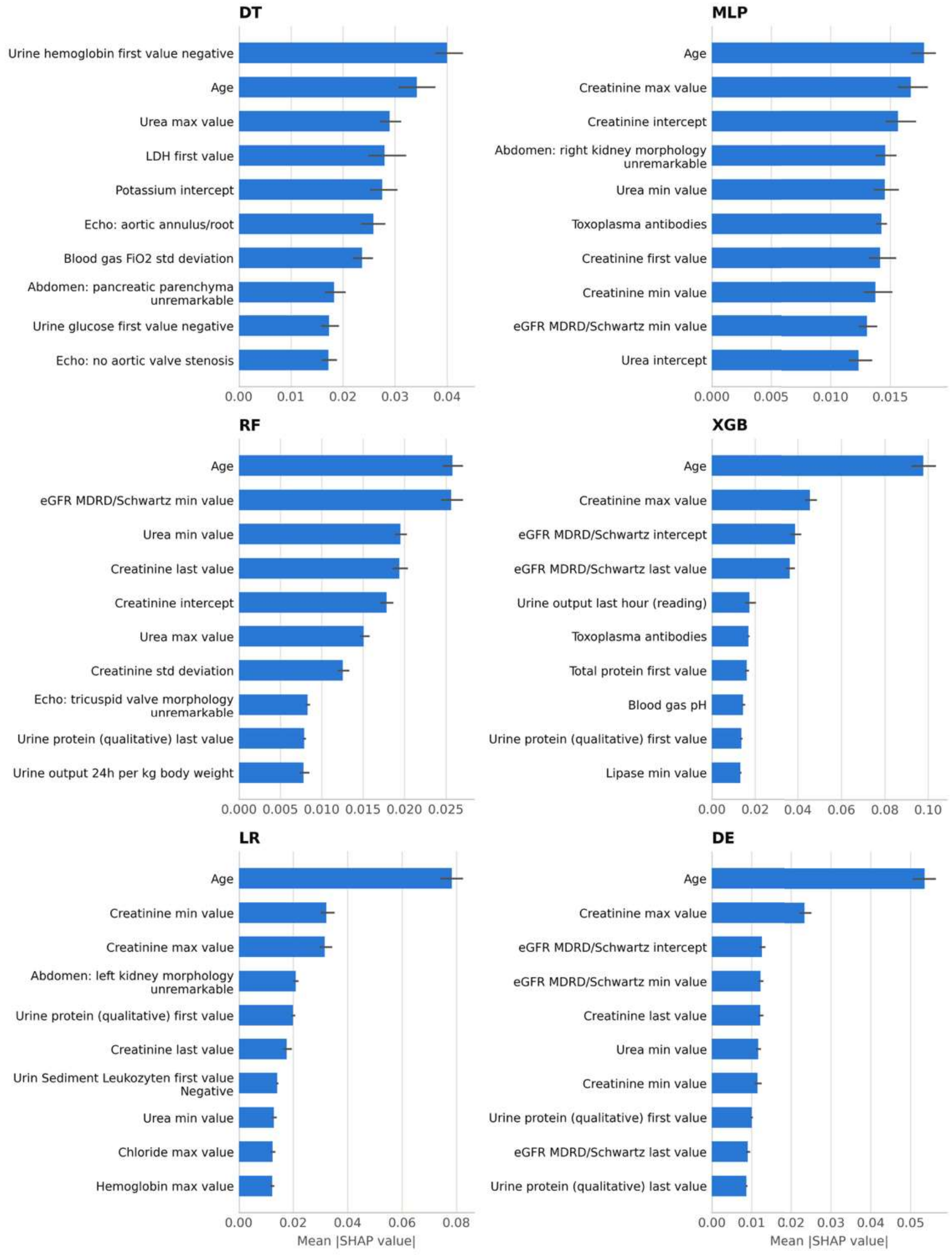

**Figure 7.** The 10 features with the largest mean absolute SHAP value per model on the test set (n = 817). Error bars show 95% BCa bootstrap confidence intervals (2,000 resamples). SHAP values refer to the predicted probability of transplantation, and those of DE are the mean of its members' SHAP values. The x-axes differ between panels. Suffixes denote aggregations of repeated measurements (first, last, minimum and maximum value, standard deviation and intercept).

## Discussion

Our results base on a concise, transferable benchmarking procedure. First, we standardize data handling and feature construction from donor data: domain-aware imputation, feature extraction from time-series, and the encoding of donor medication data, with all preprocessing parameters estimated on the training data only. Second, we perform per-model, supervised feature selection with NSGA-II to jointly optimize predictive performance (MCC) and parsimony, evaluating each candidate subset across random hyperparameter configurations to reduce overfitting to a single configuration, followed by Bayesian TPE tuning on the tailored spaces. Third, we emphasize robust, comparable evaluation: thresholds and calibrators are fitted on validation split, and all models are compared on a held-out test set that was not used for any modeling decision, with BCa bootstrap confidence intervals for the prediction performance measures. Fourth, we calibrate probabilities with Platt scaling. The post calibrator was trained on validation predictions and assess calibration with intercept, slope, and reliability plots. Finally, we unify interpretation across model families with SHAP to provide aggregated global feature importances with bootstrap confidence intervals, noting its assumptions and complementing it with model-specific diagnostics.

The DT performs worst in our experiments, which aligns with prior findings on clinical tabular data [64,65]. A single tree often lacks the capacity required for complex patterns and is vulnerable to overfitting [66]. Consistent with this, DT showed by far the largest drop from its cross-validated MCC during feature selection to its test MCC (0.430 to 0.220, compared with at most 0.069 for the other algorithms, see supplement Figure E5).

DE had the highest point estimates on all measures, in line with evidence that ensembles reduce variance and improve scoring accuracy [4,67,68]. Its advantage over LR, however, was small, and not statistically significant for any of the performance measures except the balanced Brier score A single LR thus achieved discrimination comparable to the ensemble of four models, consistent with a systematic review that found no performance benefit of ML over logistic regression for clinical prediction models [69].

Although deep learning has shown superior performance over classical ML in many domains [70,71], this advantage was not evident in our study, where LR, XGB, and RF performed comparably to the MLP. This finding likely reflects the characteristics of our dataset which is moderate in size and tabular in structure limiting the capacity of deep learning models to exploit their full representational power [30,72]. In such settings, simpler models like LR offer competitive predictive accuracy while maintaining interpretability and lower computational demands, making them attractive candidates for further clinical evaluation.

Our results also highlight the importance of metric selection. Following STRATOS, we used the AUC as primary measure of discrimination and the MCC for model development, as it incorporates all four confusion matrix components [41,42]. Like all threshold-based measures, however, the MCC is improper and depends on the threshold, which we therefore selected on validation data. Prevalence-dependent measures can be misleading, as the PR-AUC illustrates. For the five models other than DT, it ranged from 0.93 to 0.95 with transplantation and from 0.67 to 0.73 with discard as the positive class. This difference largely reflects the prevalence of each class (0.77 and 0.23), which is the PRAUC of an uninformative model, rather than a difference in model quality.

In line with prior work showing that isotonic regression require larger calibration sets than Platt scaling [73], Platt scaling calibrated better than isotonic regression on our small validation split. The need for calibration was model dependent. LR was well calibrated without it, in line with its maximum likelihood fit with the canonical logit link [85]. DT produced overconfident probabilities (slope 0.13), as expected for a single tree whose leaf estimates rest on few observations, whereas RF (slope 1.82) and DE (slope 1.26) were underconfident, consistent with the tendency of averaging to push probabilities away from 0 and 1 [73]. The MLP was only mildly miscalibrated. After calibration, all intercepts were slightly negative although the validation split and the test set had nearly the same proportion of transplantations. Calibrators fitted on a small validation split therefore transfer only partly, and models should be recalibrated in the population in which they are applied. We thus recommend applying calibration selectively, guided by model type and validated using calibration intercept, slope and reliability plots, with Platt scaling preferred for small datasets.

Finally, model interpretability remains a critical consideration for clinical adoption. Comparing model-agnostic and model-specific approaches reveals complementary strengths and limitations. Model-specific feature importances reflect algorithmic biases (e.g., regression coefficients or impurity-based scores), while SHAP offers a unified importance scale across model families [74,75]. However, SHAP attributions can be influenced by feature correlation, background data, and independence assumptions, and aggregation of absolute SHAP values omits directionality. Furthermore, SHAP values can reflect associations but cannot distinguish underlying causal mechanisms. Despite these constraints, SHAP proved valuable for identifying consistent, clinically plausible predictors, most notably age and renal function markers such as creatinine, urea and eGFR, in line with prior work [4,15,21,76]. Nevertheless, SHAP analyses should be interpreted in conjunction with model diagnostics and expert clinical judgment. In addition, the selected feature subsets agreed only moderately between the best trials of each feature selection, and the subsets of different algorithms overlapped hardly more than random subsets of the same size. The presence or absence of an individual feature among a model's top 10 should therefore not be over-interpreted.

Although, the systematic model comparison could give important insights into how well different models predict the discard of both kidneys from donor data, our study has several limitations. First, the outcome is defined at the donor level, distinguishing donors with at least one transplanted kidney from donors whose kidneys were both discarded. The models therefore identify donors whose kidneys are unlikely to be used at all, but not the discard of a single kidney, which occurred in 364 donors (8.9%) with one transplanted and one discarded kidney. A kidney-level model is expected to assign nearly identical predictions to both kidneys of a donor, as the majority of predictors are available only at the donor level. Consequently, these predictions would also be expected to closely align with those of the donor-level model. ,. Excluding these donors from the test set increased the AUC by at most 0.015 (Supplement Table E2), so they had little influence on the reported discrimination. Second, discard reflects allocation practice and acceptance decisions as well as organ quality. The models therefore learn current decision behavior rather than organ viability, and applying them could reinforce existing practice. Finally, all data came from a single organ procurement organization and were split randomly, so performance in other countries and in later years remains unknown, and external and temporal validation are needed.
Third, data entry errors may exist due to inconsistent unit systems across different German regions. Although plausibility checks (e.g., restricted ranges) take place during data entry, the difficulty of distinguishing unit-related errors from atypical values in deceased donors means that data errors may still occur. However, since these original values determined the actual graft selection, they remain valid for use in the prediction models.
Fourth, the generalizability of the clinical findings may be limited, as kidney discard processes vary across healthcare systems. However, the results of the methodological model comparison are expected to be more broadly applicable. Further studies are needed to confirm these results in different contexts.

## Conclusion

Given the abundance of models and heterogeneous clinical datasets, different ML models have been applied to predict the risk of donor discard, yet without a definitive conclusion on which model performs best [7,14,15]. Our systematic comparison using the same dataset provides more conclusive insights: An ensemble combining different model families achieves the highest point estimates, but its advantages over LR was small and not statistically significant in terms of discrimination, while it requires four models with different feature sets. Somewhat unexpectedly, no model performed substantially better than a simple linear additive model (LR), and the MLP and the tree-based ensembles (RF, XGB) reached a similar AUC and MCC. This supports the continued widespread use of the well-explainable LR model [5,9–13] and aligns with the findings of Sageshima et al. [15] who also reported similar performance between LR and other ML models. Importantly, explainability is not limited to linear models but can also be derived from more complex models using SHAP values. We found SHAP to consistently identify known clinical drivers (e.g., age, renal function markers) across model families, supporting coherent interpretation. As expected, a single DT shows clearly inferior performance, likely due to its limited capacity to capture the complexity of the data.

Our results base on a concise, transferable benchmarking procedure that prioritizes methodology over single-number outcomes. In summary, the small differences between rigorously developed models suggest that the specific choice of model type matters less for predicting donor discard than methodological rigor. External and temporal validation is needed before clinical use.

## Acknowledgments

We gratefully recognize and appreciate the support and contributions of Thomas La Rocca (DSO) and Lukas Klein, Adrian Füller, Anika Fuchs, Danilo Zähle, David Heiß, Friedrich Heitzer, Katharina Litzinger, Kevin Saliu, Leonard Holdau, and Tristan Funk (h_da) to this work.

## Funding

The research project was funded by the German Federal Ministry of Education and Research (grant 13FH019KX1). The funder had no involvement in the study design, data collection, analysis, interpretation, or the writing of the manuscript. The results presented in this article are the responsibility of the authors. The publication fees were funded by Darmstadt University of Applied Sciences.

## Code Availability

An implementation of the applied analysis methods is available on https://github.com/PeerHan/donor-kidney-discard-prediction

## Data Availability

The analyses presented in this article are based on retrospective and anonymized data from the German Organ Procurement Organization (DSO). Due to data protection and institutional restrictions, these data cannot be made publicly available.

The results derived from these data are publicly available on Zenodo (https://doi.org/10.5281/zenodo.23929751). These include the feature selection trials of each model, from which the list of all candidate features and the feature subset selected for each model can be read, all hyperparameter optimization trials, the predicted probabilities for the validation and test sets together with the observed outcomes, and the package versions of the analysis environment.

## Generative AI Use

Generative AI (Claude Opus 5.5 from Anthropic) was used during the revision of this manuscript (October 2026) to edit the language of the manuscript and of the response to the reviewers, and to write and review Python code for visualization purposes.

## Authors Contributions

Axel Rahmel, Antje Jahn, and Gunter Grieser conceived and designed the study and planned the analyses; Peer Schliephacke, Hannah Schult, Leon Mizera, and Judith Würfel designed the feature engineering and performed all analyses, with medical guidance from Carl-Ludwig Fischer-Fröhlich. Peer Schliephacke drafted the first version of the manuscript and led the revision of the manuscript in response to peer review. All authors reviewed, edited, and approved the final manuscript.

## Conflicts of Interest

The authors declare no potential conflicts of interest with respect to the research, authorship, and/or publication of this article.

## Abbreviations

AUC: Area Under the ROC Curve

BCa: Bias-corrected and accelerated Bootstrap Interval

DE: Data ensemble

DT: Decision Tree

FN: False negative

FP: False positive

LR: Logistic Regression

PPV: Positive Predictive Value

PRAUC: Area Under the Precision–Recall Curve

MCC: Matthews Correlation Coefficient

ML: Machine Learning

MLP: Multilayer Perceptron

NSGA-II: Nondominated Sorting Genetic Algorithm II

NPV: Negative Predictive Value

RF: Random Forest

SHAP: SHapley Additive exPlanations

STRATOS: STRengthening Analytical Thinking for Observational Studies

TN: True negative

TP: True positive

TPE: Tree-structured Parzan Estimator

XGB: eXtreme Gradient Boosting

**Supplementary information.**

# Appendix A Software and Hardware

We use an AMD EPYC-Milan Processor for all models trained with sklearn [1] version 1.5.2 (Decision Tree, Random Forest, Logistic Regression) and xgboost version 2.1.2. We train our MLP models with pytorch [2] version 2.5.0 and cuda version 12.2 on a NVIDIA A100 Graphic card. We use the shap package [3] version 0.46.0 for explainability.

# Appendix B Feature Engineering

## B.1 Feature Transformations

- **Age at Diabetes Diagnosis** The donor's age at the time of the diabetes diagnosis is calculated using the timestamps of their birth and the diagnosis date.
- **Duration of Diabetes** The duration of diabetes for the donor is calculated using the timestamps of the diagnosis and the time of death.
- **Duration of Alcohol Consumption** The calculation of this feature depends on whether an end is provided for the respective donor. If an end date is available, the duration of alcohol consumption (in days) is computed as the difference between the start and end dates. If no end date is provided, it is assumed that the date of hospital admission represents the end point for that donor.
- **Last Alcohol Consumption** This feature categorically represents the time elapsed since a donor's last alcohol consumption. The categorization is based on the difference between the documented end of alcohol use and the date of hospital admission. The calculation is performed as follows:
  - No alcohol consumption documented: If no start date for alcohol consumption is provided, it is assumed that the donor does not have an alcohol-related issue and is assigned to category 0.
  - Currently consuming: If a start date is available but no end date is given, it is assumed that the donor was consuming up until hospital admission. In this case, the time difference is 0 days, and the donor is assigned to the highest category, category 5.
  - Consumption ended prior to admission: If both a start and end date are available, the number of days between the end of alcohol use and hospital admission is calculated. This time span is then divided into four categories. The bins are determined dynamically based on the distribution in the training dataset, where donors who stopped drinking more recently fall into higher categories (e.g. category 4), and those whose last consumption was further in the past are placed in lower categories (e.g. category 1).
- For some categorical variables with rare values we grouped values into positive and negative to turn them into binary features, e.g. "EKG result on QRS differences" was dichotomized into "no" if no differences are reported and "yes" if "others", "MI-like", "RSB", "LSB" or "bifascicular block" is reported.

- The amount of diuresis in the last hour ($dlh$) was normalized using body weight ($bw$) (see formula B1).
- The total amount of diuresis ($dt$) within a time window ($t$) was scaled to 24 hours and normalized to body weight (see formula B2)
- We converted creatinine values to the unit mg/dL, by multiplying the values by a factor of 0.011312 (given in $\mu mol/L$ in the primary database).

$$dlh_{norm} = \frac{dlh}{bw} \tag{B1}$$

$$d_{24h/bw} = \frac{dt}{t\ (in\ hours)} \cdot \frac{24}{bw} \tag{B2}$$

# Appendix C Feature Selection

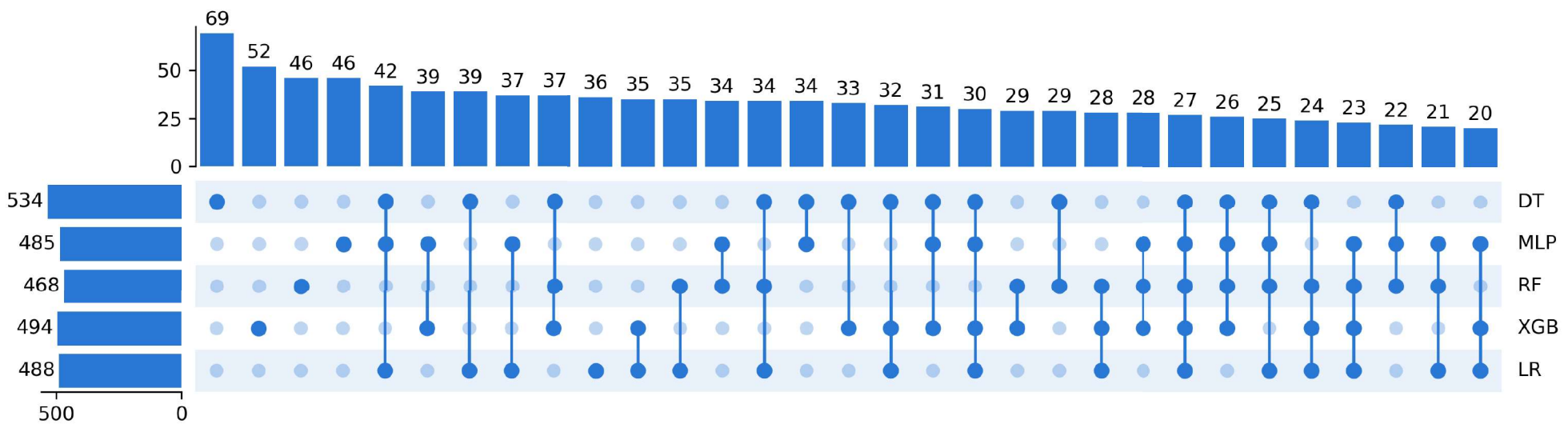

**Fig. C1** Size of intersections of the feature spaces from the feature selection per model combination.

**Table C1** Features and their meaning. ICD-10 codes follow the German modification (ICD-10-GM).

| **Feature** | **Meaning** |
|---|---|
| Abdomen: free fluid pronounced | Abdominal imaging: pronounced free intra-abdominal fluid |
| Age | Age in years |
| Albumin min value | Lowest recorded albumin value |
| Alcohol ICD-10 F10.0 | Alcohol use coded as ICD-10 F10.0: mental and behavioural disorders due to use of alcohol, acute intoxication |
| ALT/SGPT last value | Last recorded alanine aminotransferase (ALT, also SGPT) |
| aPTT value count | Number of activated partial thromboplastin time (aPTT) measurements |
| Blood gas base excess max value | Highest base excess in blood gas analysis |
| Blood gas $FiO_2$ first value | First recorded fraction of inspired oxygen ($FiO_2$) in blood gas analysis |
| BMI | Body mass index |
| Bronchoscopy: tracheal mucosa | Bronchoscopic finding of the tracheal mucosa |
| Coronary angiography: mid LAD | Coronary angiography finding of the middle segment of the left anterior descending artery (LAD; German: RIVA) |
| Coronary angiography: RCA and branches | Coronary angiography finding of the right coronary artery (RCA) and its branches |
| Diabetes ICD-10 E11.50 | Diabetes coded as ICD-10 E11.50: type 2 diabetes mellitus with peripheral circulatory complications, not stated as uncontrolled |
| Drug use grading: formerly intravenous | Grading of drug use: former intravenous drug use |
| Drug use ICD-10 F12.0 | Drug use coded as ICD-10 F12.0: mental and behavioural disorders due to use of cannabinoids, acute intoxication |
| Echo: posterior mitral leaflet morphology unremarkable | Echocardiography: unremarkable morphology of the posterior mitral valve leaflet |
| HBsAg | Hepatitis B surface antigen (serology) |
| HSD ICD-10 S06.9 | HSD diagnosis coded as ICD-10 S06.9: intracranial injury, unspecified |
| LDH min value | Lowest recorded lactate dehydrogenase (LDH) |
| Leukocytes std deviation | Standard deviation of recorded leukocyte (white blood cell) counts |
| Malignancy ICD-10 C43.9 | Malignancy coded as ICD-10 C43.9: malignant melanoma of skin, unspecified |
| Malignancy ICD-10 C44.0 | Malignancy coded as ICD-10 C44.0: other malignant neoplasm of skin, skin of lip |
| Malignancy ICD-10 C80.9 | Malignancy coded as ICD-10 C80.9: malignant neoplasm, primary site unspecified |
| Malignancy ICD-10 D07.3 | Malignancy coded as ICD-10 D07.3: carcinoma in situ of other and unspecified female genital organs |
| Malignancy ICD-10 D42.9 | Malignancy coded as ICD-10 D42.9: neoplasm of uncertain or unknown behaviour of the meninges, unspecified |
| Urine protein (qualitative) last value | Last result of the qualitative urine protein test |
| Urine sediment leukocytes first value negative | First urine sediment result for leukocytes: negative |

# Appendix D Hyperparameter Optimization

## D.1 Search Spaces

**Table D1** Hyperparameter search space for DT (100 trials; best MCC = 0.387 in trial 67).

| Hyperparameter | Type | Search space | Selected |
|---|---|---|---|
| class_weight | categorical | {None, balanced} | None |
| max_depth | integer | 1–49 | 10 |
| max_features | categorical | {sqrt, log2} | sqrt |
| min_samples_leaf | integer | 1–20 | 20 |
| min_samples_split | integer | 2–20 | 12 |

**Table D2** Hyperparameter search space for MLP (100 trials; best MCC = 0.522 in trial 83).

| Hyperparameter | Type | Search space | Selected |
|---|---|---|---|
| activation_function | categorical | {Mish, Swish, Leaky, ELU, ReLU, GELU} | GELU |
| batchnorm | integer | 0–1 | 0 |
| class_weights | integer | 0–1 | 0 |
| dropout | float | 0–0.5 | 0.45 |
| hidden_dim | integer | 131–1488 | 746 |
| init_lr | float | 0.000107–0.0711 | 0.000952 |
| n_layer | integer | 2–15 | 3 |
| weight_decay | float | 1.17e-10–7.3e-07 | 1.28e-09 |

**Table D3** Hyperparameter search space for RF (100 trials; best MCC = 0.551 in trial 73).

| Hyperparameter | Type | Search space | Selected |
|---|---|---|---|
| class_weight | categorical | {balanced, None} | balanced |
| max_depth | integer | 2–50 | 28 |
| max_features | categorical | {sqrt, log2} | sqrt |
| min_samples_leaf | integer | 1–20 | 6 |
| min_samples_split | integer | 2–20 | 14 |
| n_estimators | integer | 12–499 | 474 |

**Table D4** Hyperparameter search space for XGB (100 trials; best MCC = 0.578 in trial 16).

| Hyperparameter | Type | Search space | Selected |
|---|---|---|---|
| colsample_bytree | float | 0.755–1 | 0.997 |
| early_stopping_rounds | integer | 5–100 | 50 |
| learning_rate | float | 0.00117–0.0983 | 0.00726 |
| max_depth | integer | 2–15 | 5 |
| min_child_weight | integer | 1–20 | 4 |
| n_estimators | integer | 200–1500 | 1400 |
| reg_alpha | float | 0.00108–12.1 | 0.0627 |
| reg_lambda | float | 0.00118–18.4 | 0.0526 |
| subsample | float | 0.762–0.999 | 0.945 |
| tree_method | categorical | {hist, approx} | approx |
| use_class_weights | integer | 0–1 | 1 |

**Table D5** Hyperparameter search space for LR (100 trials; best MCC = 0.553 in trial 65).

| Hyperparameter | Type | Search space | Selected |
|---|---|---|---|
| C | float | 0.00121–9.45 | 0.0409 |
| class_weight | categorical | {None, balanced} | None |
| l1_ratio | float | 0.0278–0.992 | 0.632 |

## D.2 Best Configs

**Table D6** The 5 best of 100 completed hyperparameter optimization trials for LR, ranked by MCC; the first column is the selected configuration.

| Hyperparameter | Trial 65 | Trial 37 | Trial 15 | Trial 71 | Trial 63 |
|---|---|---|---|---|---|
| MCC | 0.553 | 0.552 | 0.551 | 0.550 | 0.550 |
| C | 0.0409 | 0.0369 | 0.0371 | 0.0329 | 0.0415 |
| class_weight | None | None | None | None | None |
| l1_ratio | 0.632 | 0.574 | 0.632 | 0.509 | 0.589 |

**Table D7** The 5 best of 100 completed hyperparameter optimization trials for DT, ranked by MCC; the first column is the selected configuration.

| Hyperparameter | Trial 67 | Trial 68 | Trial 83 | Trial 92 | Trial 69 |
|---|---|---|---|---|---|
| MCC | 0.387 | 0.387 | 0.387 | 0.384 | 0.379 |
| class_weight | None | None | None | None | None |
| max_depth | 10 | 10 | 10 | 5 | 10 |
| max_features | sqrt | sqrt | sqrt | sqrt | sqrt |
| min_samples_leaf | 20 | 20 | 20 | 20 | 19 |
| min_samples_split | 12 | 14 | 18 | 13 | 13 |

**Table D8** The 5 best of 100 completed hyperparameter optimization trials for RF, ranked by MCC; the first column is the selected configuration.

| Hyperparameter | Trial 73 | Trial 93 | Trial 74 | Trial 86 | Trial 64 |
|---|---|---|---|---|---|
| MCC | 0.551 | 0.547 | 0.547 | 0.546 | 0.544 |
| class_weight | balanced | balanced | balanced | balanced | balanced |
| max_depth | 28 | 14 | 26 | 25 | 48 |
| max_features | sqrt | sqrt | sqrt | sqrt | sqrt |
| min_samples_leaf | 6 | 2 | 7 | 8 | 4 |
| min_samples_split | 14 | 20 | 17 | 17 | 13 |
| n_estimators | 474 | 496 | 462 | 454 | 442 |

**Table D9** The 5 best of 100 completed hyperparameter optimization trials for XGB, ranked by MCC; the first column is the selected configuration.

| Hyperparameter | Trial 16 | Trial 55 | Trial 43 | Trial 58 | Trial 19 |
|---|---|---|---|---|---|
| MCC | 0.578 | 0.574 | 0.574 | 0.573 | 0.572 |
| colsample_bytree | 0.997 | 0.992 | 0.933 | 0.95 | 0.996 |
| early_stopping_rounds | 50 | 85 | 70 | 95 | 65 |
| learning_rate | 0.00726 | 0.00944 | 0.01 | 0.0186 | 0.00693 |
| max_depth | 5 | 5 | 4 | 9 | 4 |
| min_child_weight | 4 | 1 | 1 | 7 | 1 |
| n_estimators | 1400 | 1250 | 1200 | 1050 | 1250 |
| reg_alpha | 0.0627 | 0.00295 | 0.00567 | 0.58 | 0.00372 |
| reg_lambda | 0.0526 | 0.00236 | 0.0476 | 0.0135 | 0.0297 |
| subsample | 0.945 | 0.992 | 0.916 | 0.997 | 0.943 |
| tree_method | approx | hist | hist | approx | approx |
| use_class_weights | 1 | 1 | 1 | 1 | 1 |

**Table D10** The 5 best of 100 completed hyperparameter optimization trials for MLP, ranked by MCC; the first column is the selected configuration.

| Hyperparameter | Trial 83 | Trial 3 | Trial 10 | Trial 36 | Trial 15 |
|---|---|---|---|---|---|
| MCC | 0.522 | 0.512 | 0.512 | 0.511 | 0.509 |
| activation_function | GELU | Leaky | Leaky | Leaky | Leaky |
| batchnorm | 0 | 1 | 1 | 0 | 0 |
| class_weights | 0 | 0 | 0 | 0 | 0 |
| dropout | 0.45 | 0.5 | 0.4 | 0.5 | 0.5 |
| hidden_dim | 746 | 1373 | 1363 | 724 | 1175 |
| init_lr | 0.000952 | 0.000232 | 0.000291 | 0.0017 | 0.000205 |
| n_layer | 3 | 2 | 2 | 3 | 4 |
| weight_decay | 1.28e-09 | 5.76e-09 | 1.77e-07 | 1.3e-09 | 6.18e-09 |

## D.3 Feasability

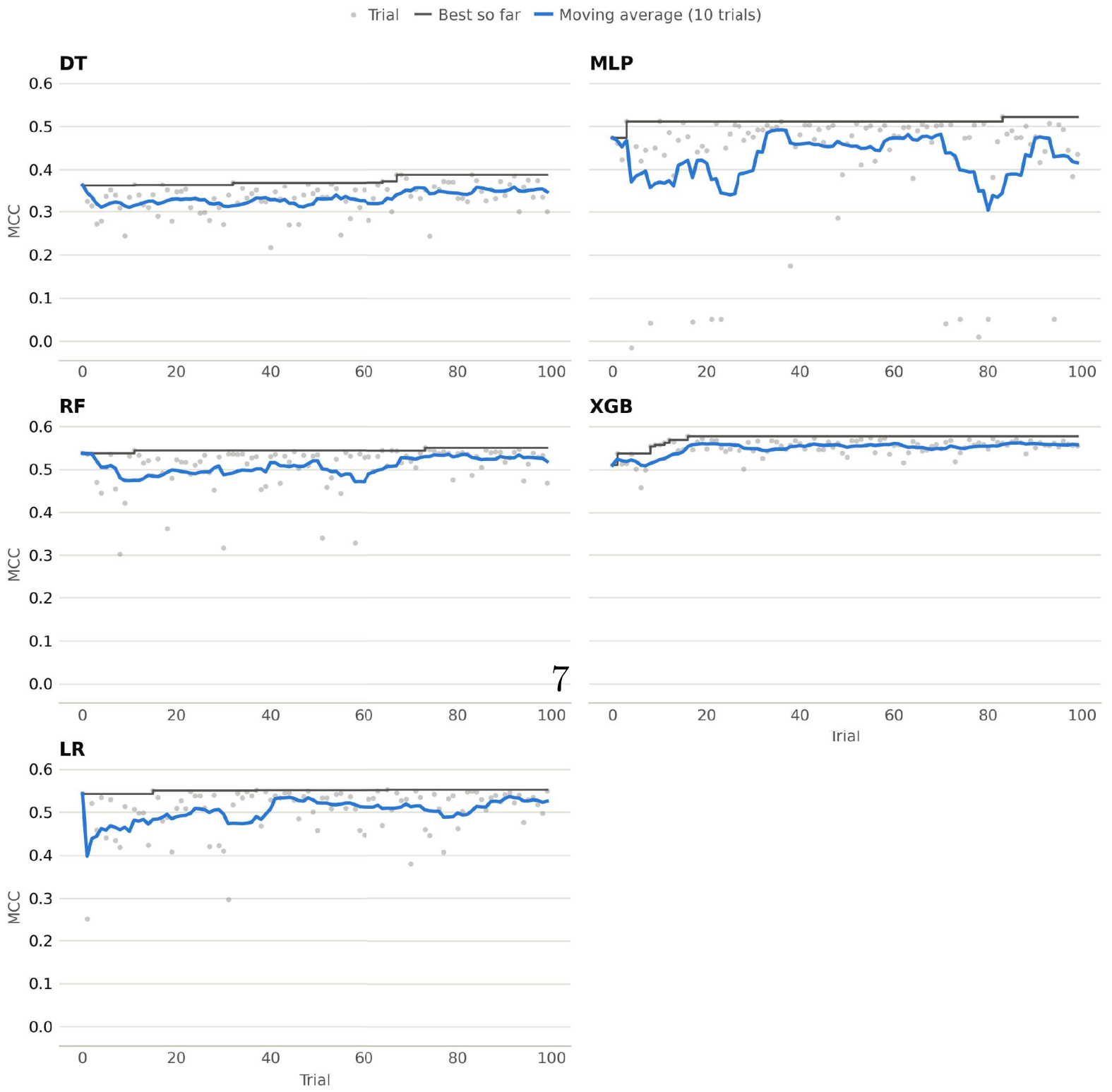

**Fig. D1** Hyperparameter optimization history. For each model, gray points show the MCC of each of the 100 trials (5-fold cross-validation), the dark gray line the best MCC reached so far, and the blue line the moving average over 10 trials.

# Appendix E Performance Evaluation

## E.1 Threshold Analysis

**Table E1** Classification thresholds selected on the validation split (n = 490; Transplanted: 379; rule: share of predicted Transplanted equal to its prevalence). Thresholds refer to the predicted probability of Transplanted; Transplanted is predicted if it exceeds the threshold. Sensitivity and specificity refer to Transplanted as the positive class. The rule threshold was used only if it increased the MCC on the validation data; otherwise the default of 0.5 was kept (Selected). Differences are rule threshold minus 0.5 on the validation data, with 95% BCa bootstrap confidence intervals (2000 resamples).

| Model | Rule threshold | Selected | Metric | At 0.5 | At rule threshold | Difference (95% CI) |
|---|---|---|---|---|---|---|
| DT | 0.603 | 0.500 | MCC | 0.308 | 0.286 | −0.022 (−0.084 to 0.050) |
| | | | Sensitivity | 0.910 | 0.836 | −0.074 (−0.104 to −0.050) |
| | | | Specificity | 0.351 | 0.450 | 0.099 (0.053 to 0.165) |
| RF | 0.526 | 0.500 | MCC | 0.647 | 0.639 | −0.008 (−0.051 to 0.039) |
| | | | Sensitivity | 0.945 | 0.918 | −0.026 (−0.047 to −0.013) |
| | | | Specificity | 0.667 | 0.721 | 0.054 (0.019 to 0.117) |
| MLP | 0.583 | 0.500 | MCC | 0.551 | 0.546 | −0.006 (−0.048 to 0.045) |
| | | | Sensitivity | 0.929 | 0.897 | −0.032 (−0.053 to −0.018) |
| | | | Specificity | 0.586 | 0.649 | 0.063 (0.027 to 0.126) |
| XGB | 0.460 | 0.460 | MCC | 0.605 | 0.627 | 0.023 (0.009 to 0.052) |
| | | | Sensitivity | 0.902 | 0.916 | 0.013 (0.005 to 0.030) |
| | | | Specificity | 0.712 | 0.712 | 0.000 (0.000 to 0.000) |
| LR | 0.619 | 0.619 | MCC | 0.569 | 0.569 | 0.000 (−0.055 to 0.059) |
| | | | Sensitivity | 0.945 | 0.902 | −0.042 (−0.065 to −0.025) |
| | | | Specificity | 0.568 | 0.667 | 0.099 (0.053 to 0.164) |
| DE | 0.552 | 0.500 | MCC | 0.660 | 0.616 | −0.044 (−0.088 to −0.005) |
| | | | Sensitivity | 0.947 | 0.913 | −0.034 (−0.059 to −0.021) |
| | | | Specificity | 0.676 | 0.703 | 0.027 (0.009 to 0.082) |

## E.2 Further Metrics

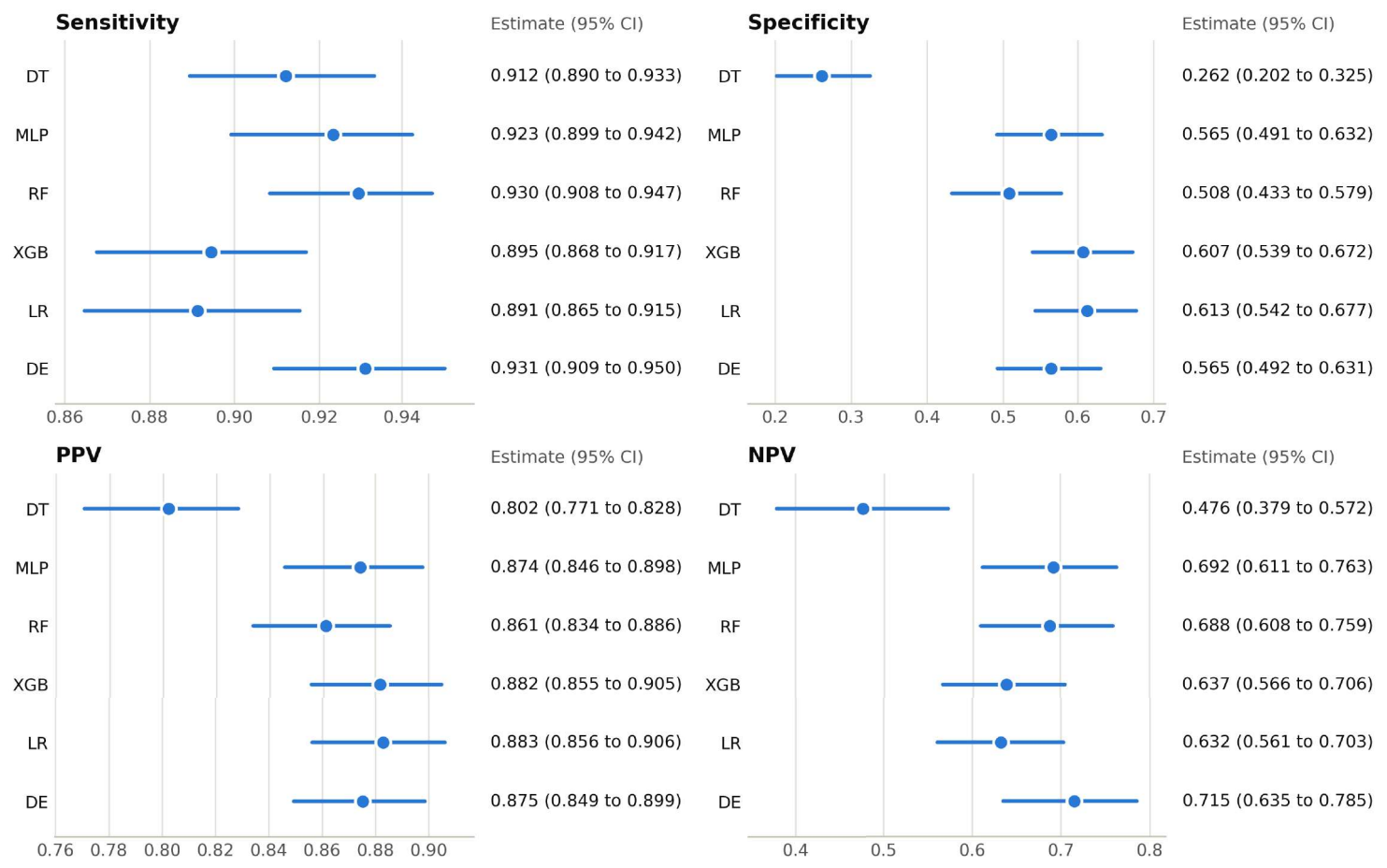

**Fig. E1** Sensitivity, specificity, positive predictive value (PPV), and negative predictive value (NPV) on the test set, with transplanted as the positive class. Points show estimates and horizontal lines 95% bootstrap CIs from 2000 resamples.

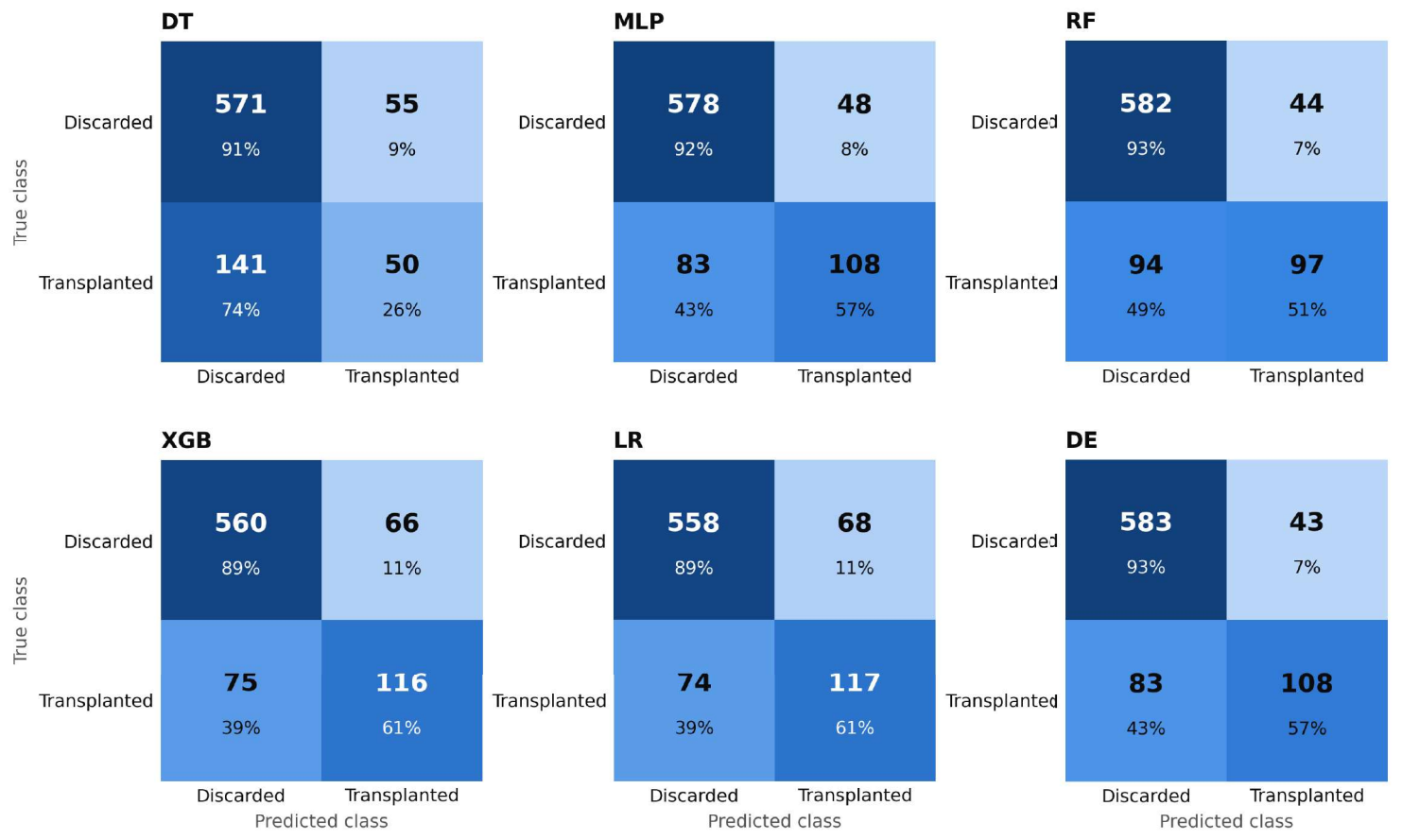

**Fig. E2** Confusion matrices of all models on the test set ($n = 817$). Rows show the true class and columns the predicted class. Cells give the number of cases and the percentage of the respective true class.

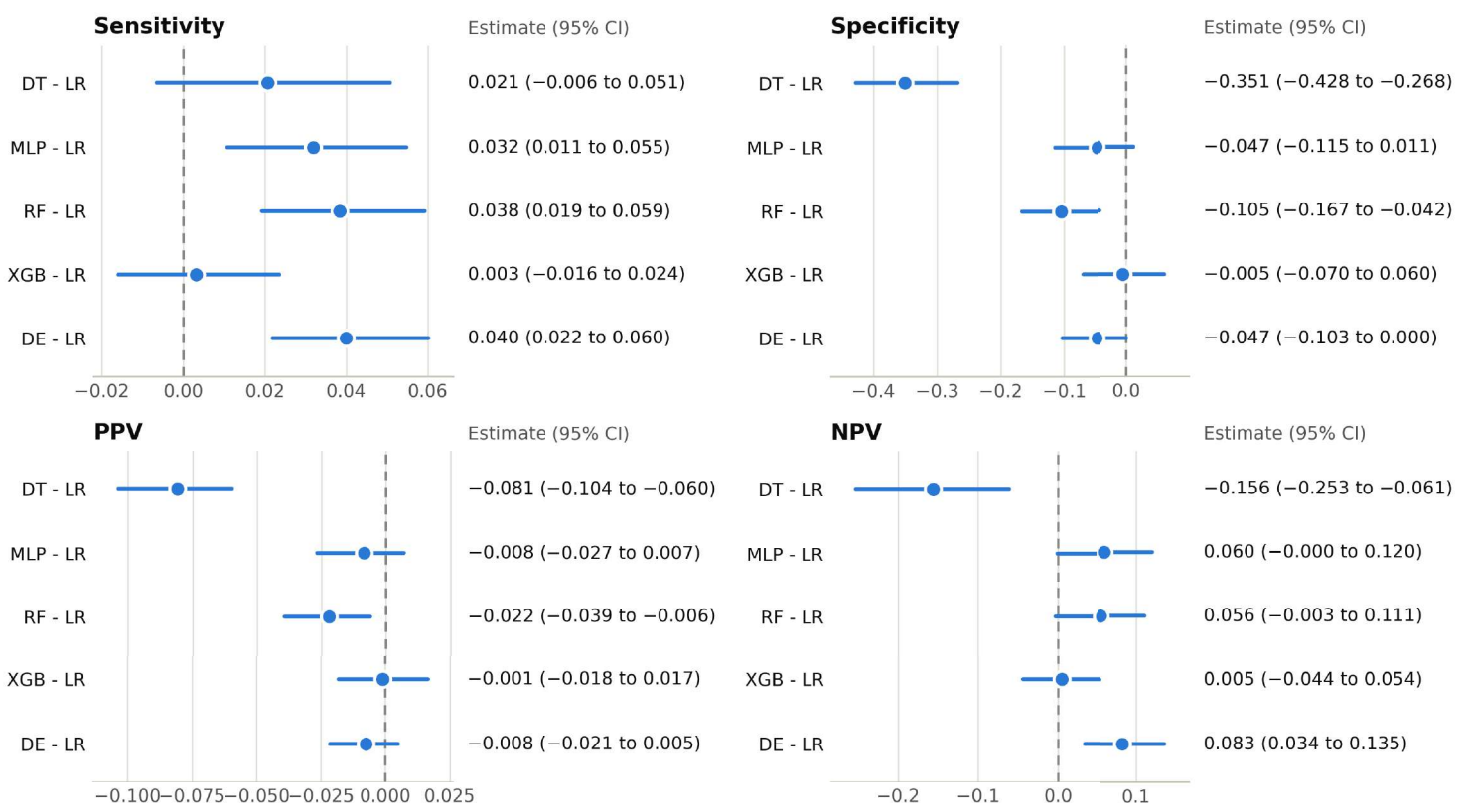

**Fig. E3** Sensitivity, specificity, positive predictive value (PPV), and negative predictive value (NPV) on the test set compared to the LR model as reference. Transplanted is the positive class. Points show estimates and horizontal lines 95% bootstrap CIs from 2000 resamples.

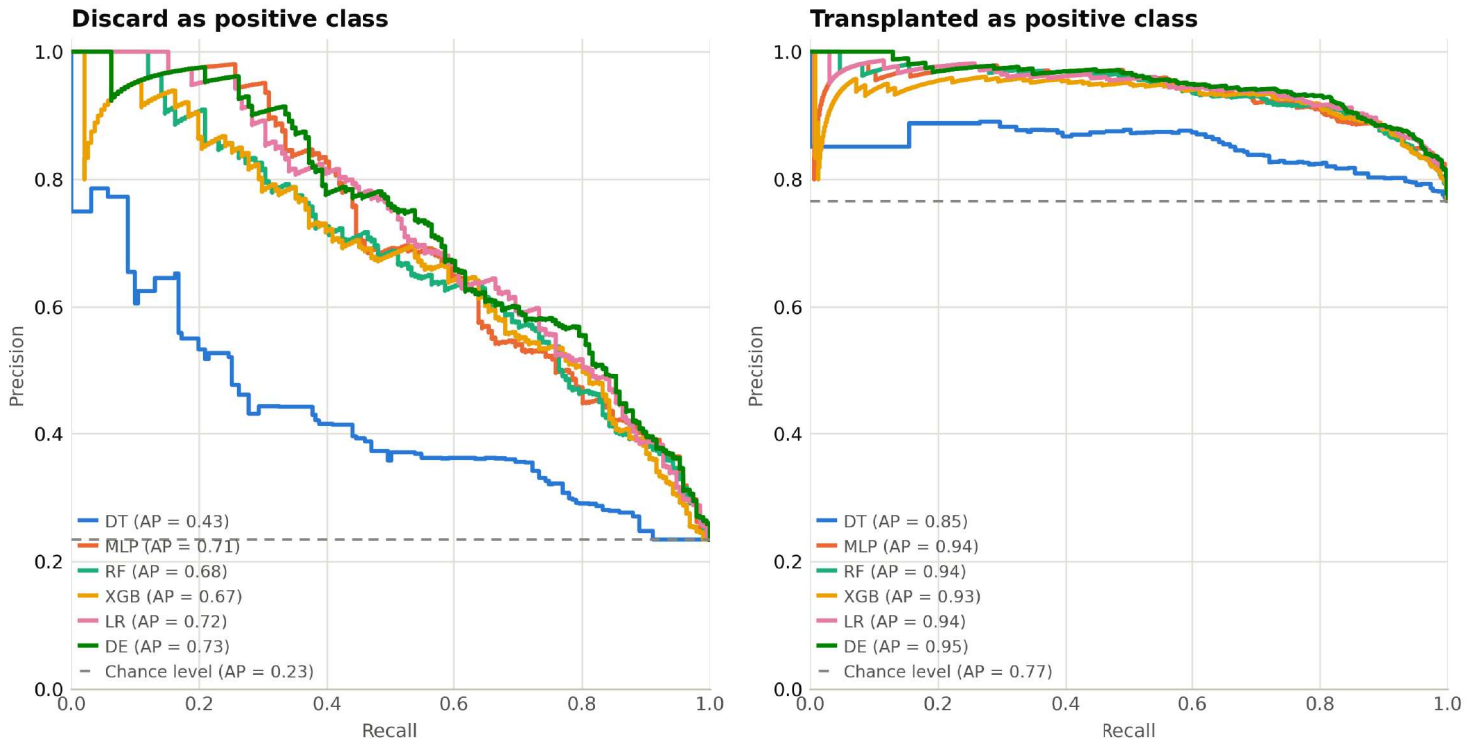

**Fig. E4** Precision–recall curves on the test set (n = 817) with discard (left) and transplantation of at least one kidney (right) as the positive class. The legend gives the average precision (PR-AUC) of each model. The dashed line marks the chance level, which equals the prevalence of the positive class.

## E.3 Sensitivity Analysis

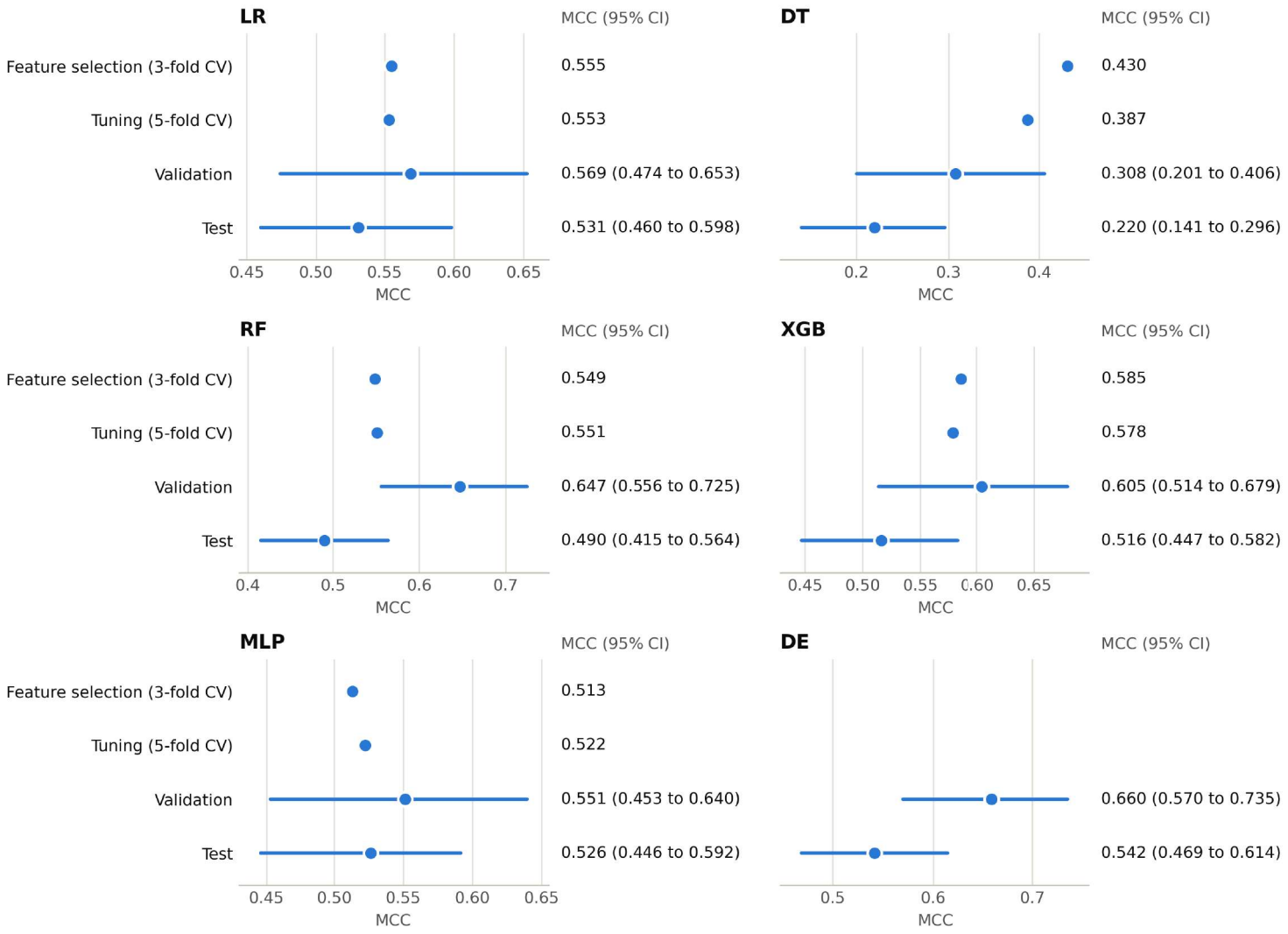

**Fig. E5** MCC of each model across the stages of model development: feature selection (3-fold cross-validation), hyperparameter tuning (5-fold cross-validation), validation set, and test set. Cross-validation results are shown as point estimates; for the validation and test sets, horizontal lines show 95% bootstrap CIs. For DE, only validation and test results are available.

**Table E2** Sensitivity analysis without the 70 test donors with one transplanted and one discarded kidney. Estimates with 95% BCa bootstrap confidence intervals (2,000 resamples) at the thresholds selected on the validation split. The prevalence of transplantation is 76.6% in the full test set and 74.4% without discordant donors. Intercept and slope refer to the uncalibrated predicted probability that at least one kidney is transplanted.

| Metric | Model | Full test set (n = 817) | Without discordant donors (n = 747) |
|---|---|---|---|
| AUC | DT | 0.684 (0.636 to 0.727) | 0.684 (0.636 to 0.725) |
| | MLP | 0.853 (0.813 to 0.881) | 0.865 (0.830 to 0.890) |
| | RF | 0.849 (0.814 to 0.877) | 0.863 (0.829 to 0.889) |
| | XGB | 0.842 (0.805 to 0.872) | 0.857 (0.820 to 0.886) |
| | LR | 0.861 (0.825 to 0.890) | 0.875 (0.842 to 0.902) |
| | DE | 0.870 (0.834 to 0.896) | 0.882 (0.851 to 0.907) |
| MCC | DT | 0.220 (0.141 to 0.296) | 0.220 (0.142 to 0.298) |
| | MLP | 0.526 (0.446 to 0.592) | 0.540 (0.459 to 0.603) |
| | RF | 0.490 (0.415 to 0.564) | 0.512 (0.434 to 0.580) |
| | XGB | 0.511 (0.442 to 0.579) | 0.549 (0.474 to 0.614) |
| | LR | 0.510 (0.436 to 0.578) | 0.556 (0.485 to 0.622) |
| | DE | 0.542 (0.469 to 0.614) | 0.568 (0.487 to 0.629) |
| PR-AUC (Discard) | DT | 0.425 (0.350 to 0.492) | 0.448 (0.378 to 0.515) |
| | MLP | 0.706 (0.633 to 0.759) | 0.744 (0.684 to 0.790) |
| | RF | 0.681 (0.611 to 0.738) | 0.726 (0.663 to 0.779) |
| | XGB | 0.671 (0.595 to 0.732) | 0.739 (0.676 to 0.790) |
| | LR | 0.721 (0.656 to 0.772) | 0.765 (0.705 to 0.812) |
| | DE | 0.728 (0.658 to 0.779) | 0.774 (0.716 to 0.819) |
| Intercept | DT | −0.07 (−0.32 to 0.16) | −0.22 (−0.46 to 0.03) |
| | MLP | 0.19 (−0.02 to 0.41) | −0.02 (−0.25 to 0.19) |
| | RF | 0.48 (0.33 to 0.64) | 0.33 (0.17 to 0.49) |
| | XGB | 0.54 (0.31 to 0.77) | 0.29 (0.06 to 0.52) |
| | LR | −0.01 (−0.21 to 0.18) | −0.22 (−0.43 to −0.03) |
| | DE | 0.32 (0.14 to 0.51) | 0.14 (−0.04 to 0.31) |
| Slope | DT | 0.13 (0.07 to 0.21) | 0.12 (0.06 to 0.20) |
| | MLP | 0.83 (0.68 to 0.96) | 0.86 (0.72 to 0.98) |
| | RF | 1.82 (1.55 to 2.10) | 1.93 (1.65 to 2.23) |
| | XGB | 0.79 (0.67 to 0.92) | 0.88 (0.75 to 1.01) |
| | LR | 1.02 (0.86 to 1.18) | 1.09 (0.93 to 1.28) |
| | DE | 1.26 (1.08 to 1.46) | 1.35 (1.15 to 1.54) |

# Appendix F Calibration

## F.1 Balanced Brier Scores

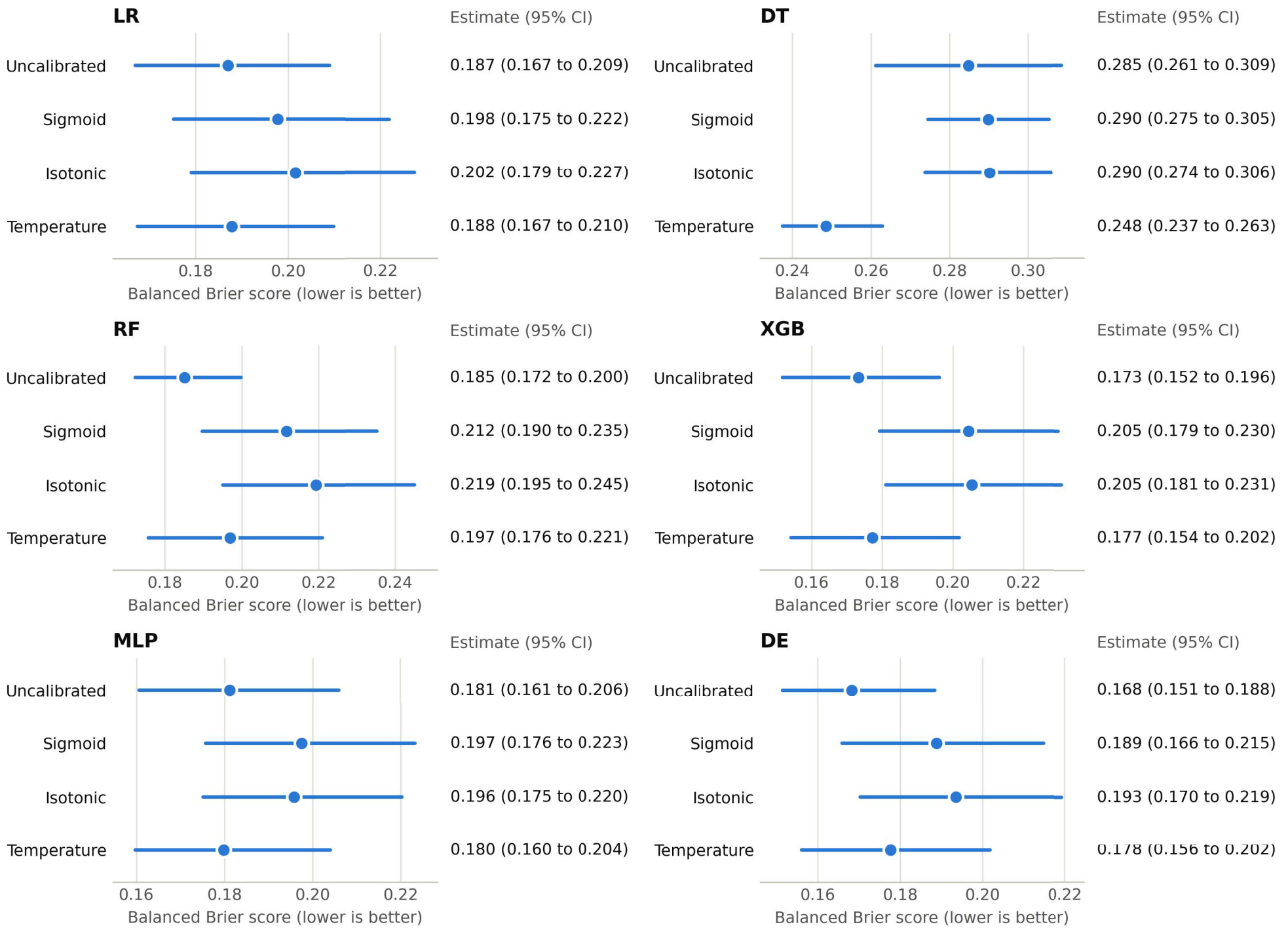

**Fig. F1** Balanced Brier score on the test set for each model without calibration and after sigmoid (Platt) scaling, isotonic regression, and temperature scaling (fitted on the validation set). The balanced Brier score averages the Brier scores of the two classes and thus weights them equally. Points show estimates and horizontal lines 95% bootstrap CIs; lower values are better.

## F.2 Brier Scores

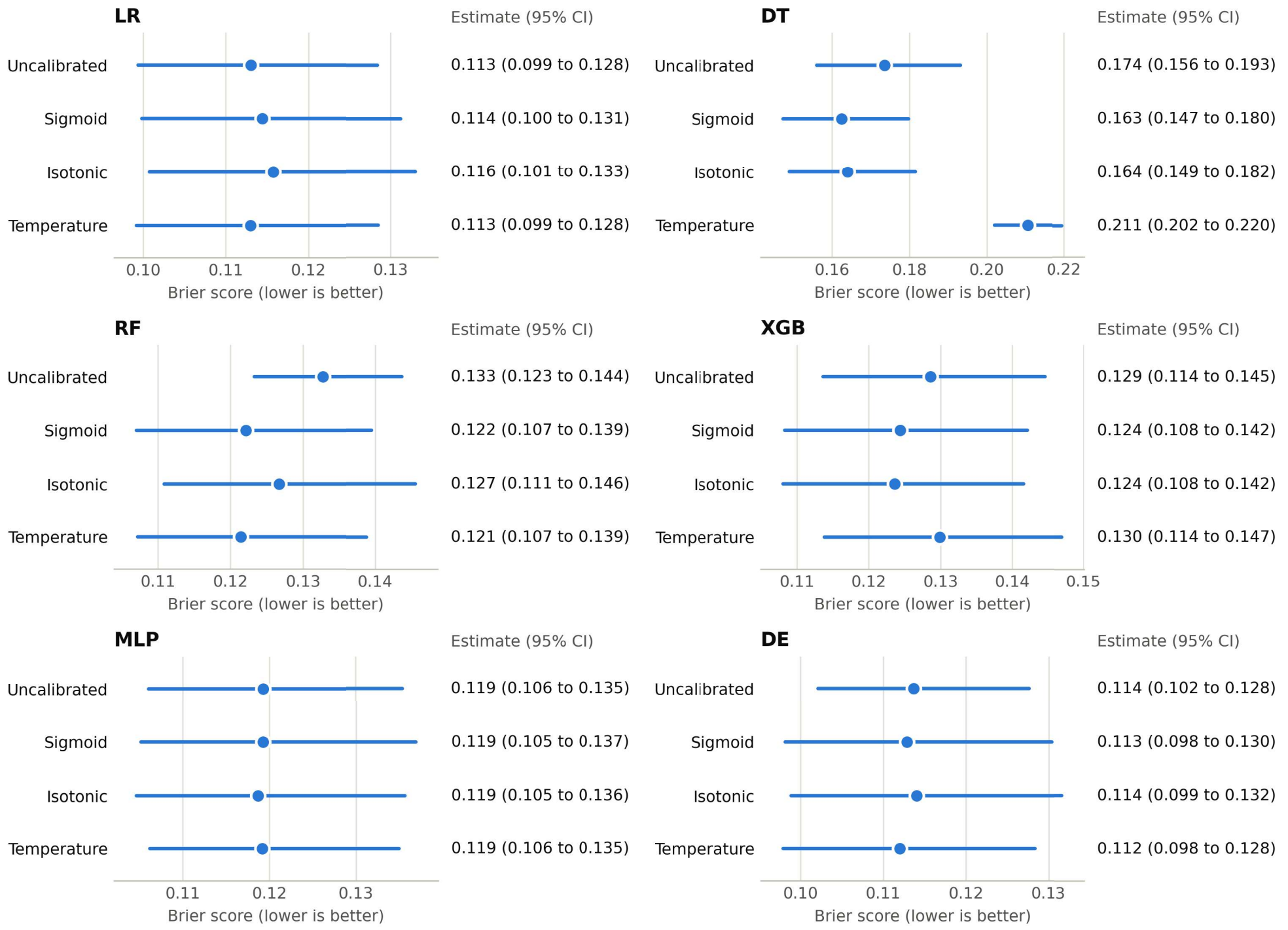

**Fig. F2** Brier score on the test set for each model without calibration and after sigmoid (Platt) scaling, isotonic regression, and temperature scaling (fitted on the validation set). Points show estimates and horizontal lines 95% bootstrap CIs; lower values are better.

## F.3 Post-Calibration

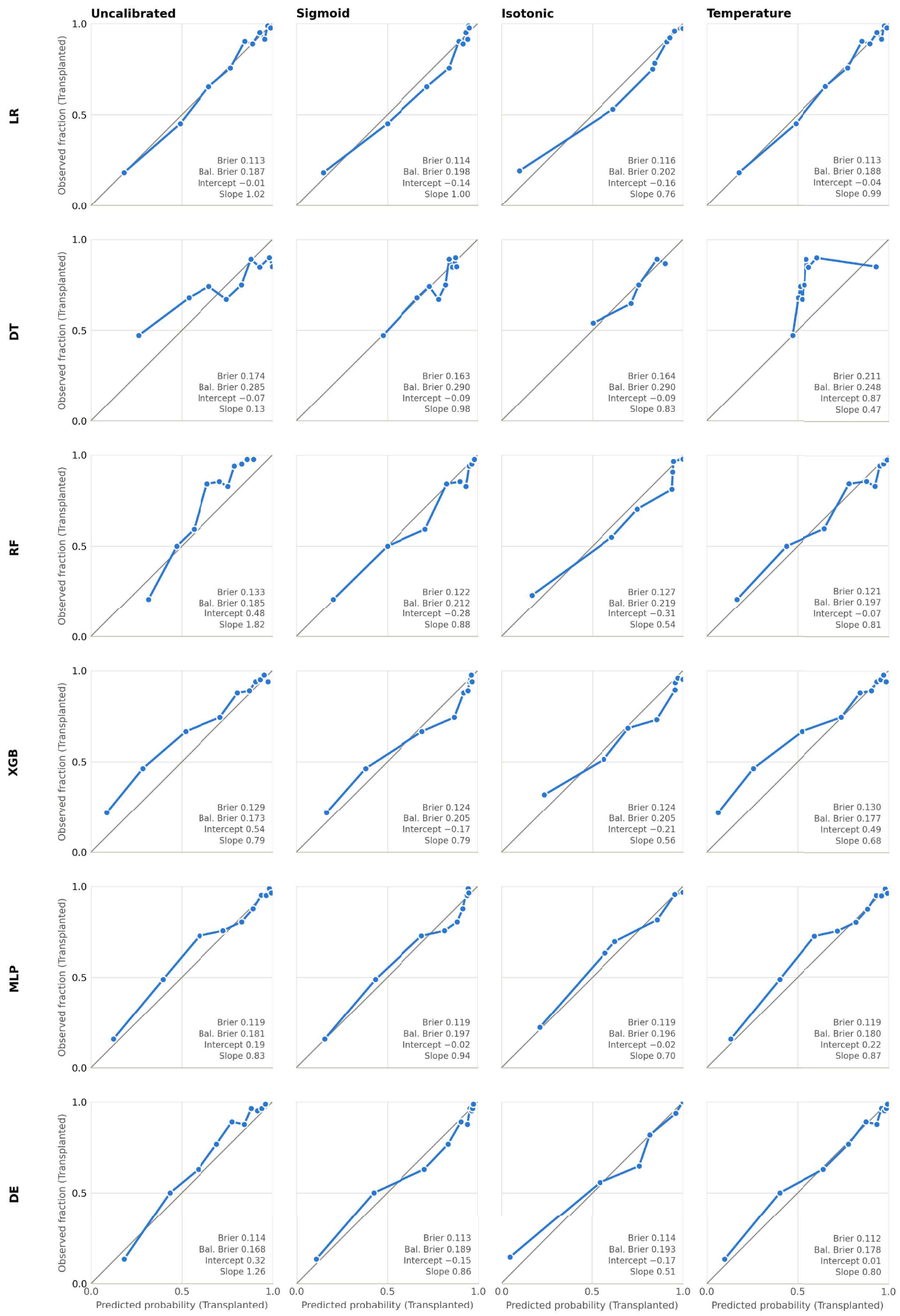

**Fig. F3** Calibration curves on the test set for each model (rows) without calibration and after sigmoid (Platt) scaling, isotonic regression, and temperature scaling (columns). Alle post-calibration methods were fitted on the validation set. Points show the observed fraction of transplanted cases against the predicted probability per bin; the diagonal indicates perfect calibration. Insets give the Brier score, balanced Brier score, calibration intercept (ideal 0), and calibration slope (ideal 1).

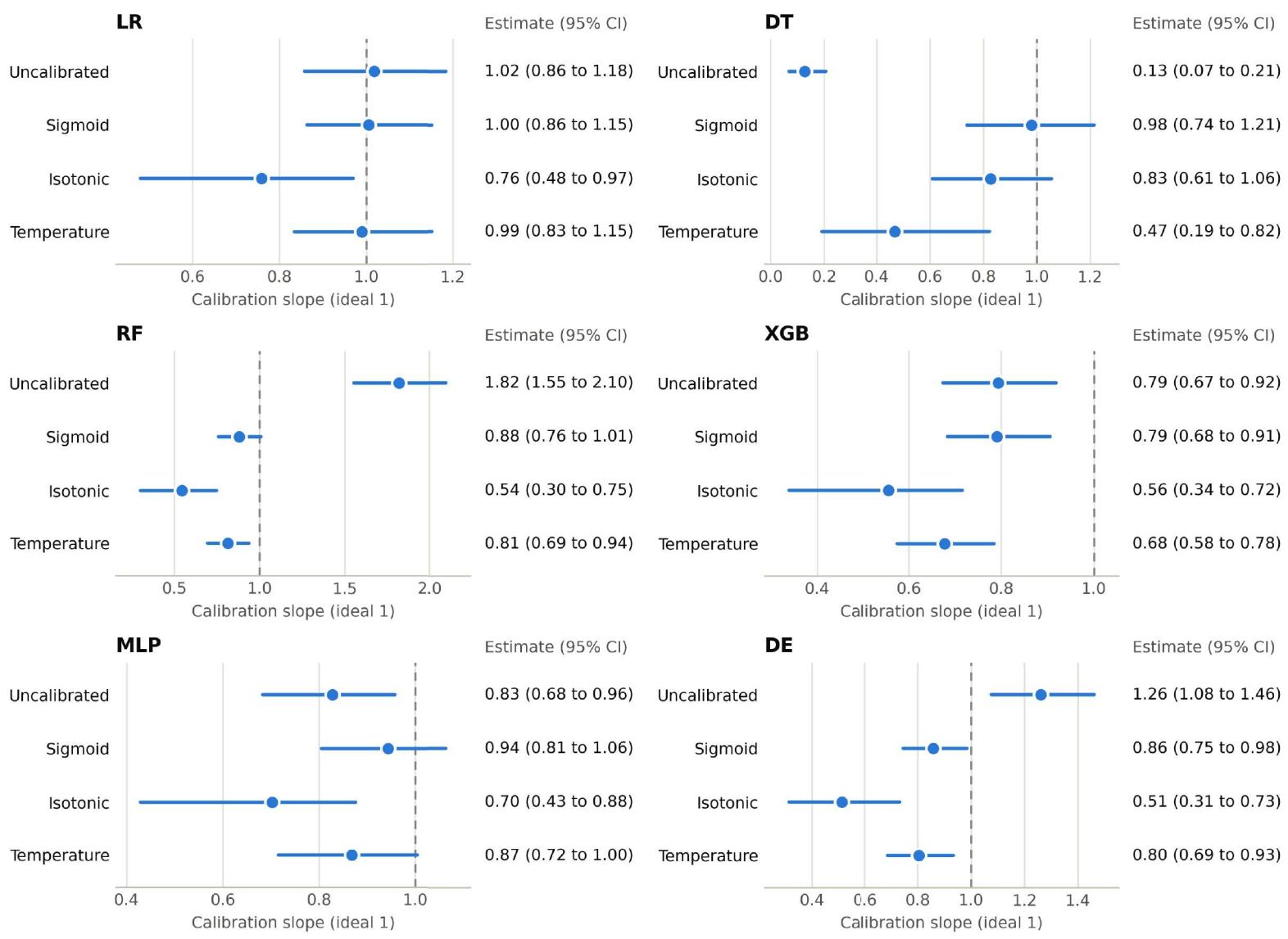

**Fig. F4** Calibration slope of all models on the test set (n = 817) without and after post-hoc calibration, with 95% BCa bootstrap confidence intervals (2,000 resamples). The dashed lines mark the ideal values of 0. The positive class is the predicted probability that at least one kidney is transplanted.

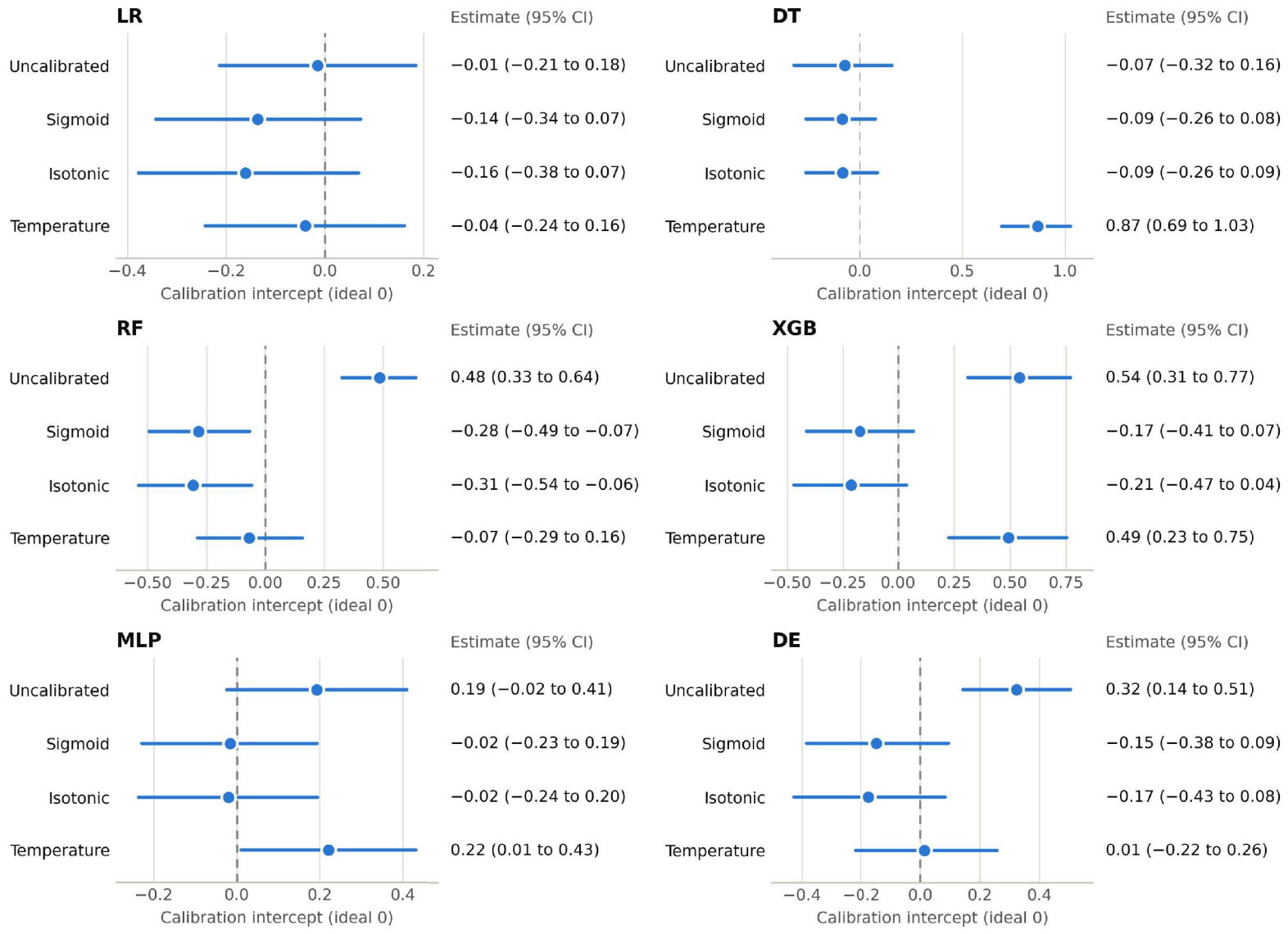

**Fig. F5** Calibration intercept of all models on the test set (n = 817) without and after post-hoc calibration, with 95% BCa bootstrap confidence intervals (2,000 resamples). The dashed lines mark the ideal values of 1. The positive class is the predicted probability that at least one kidney is transplanted.

# Appendix G Explainability

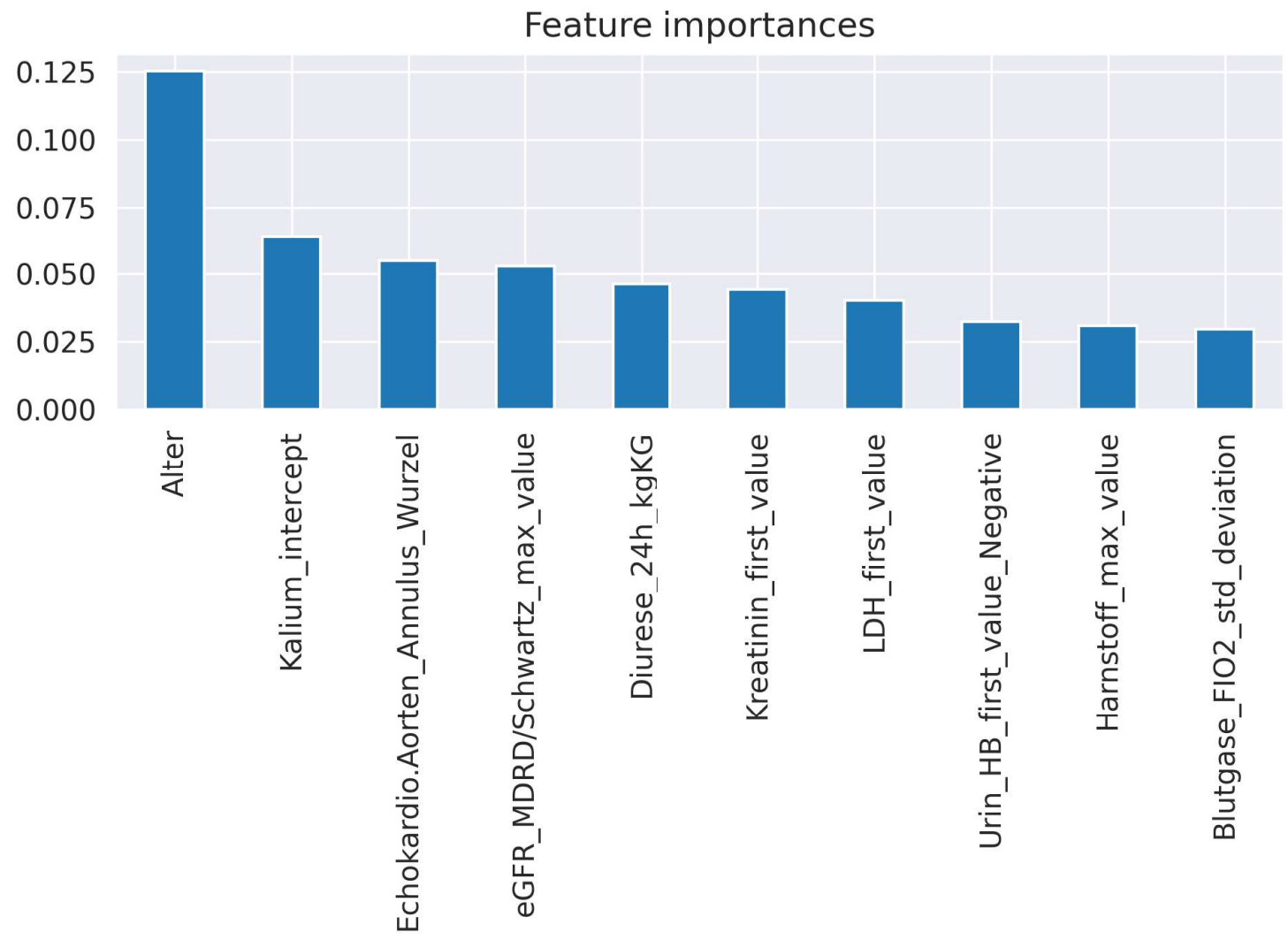

**Fig. G1** Model specific feature importance of DT (mean decrease in impurity) for the 10 most important features.

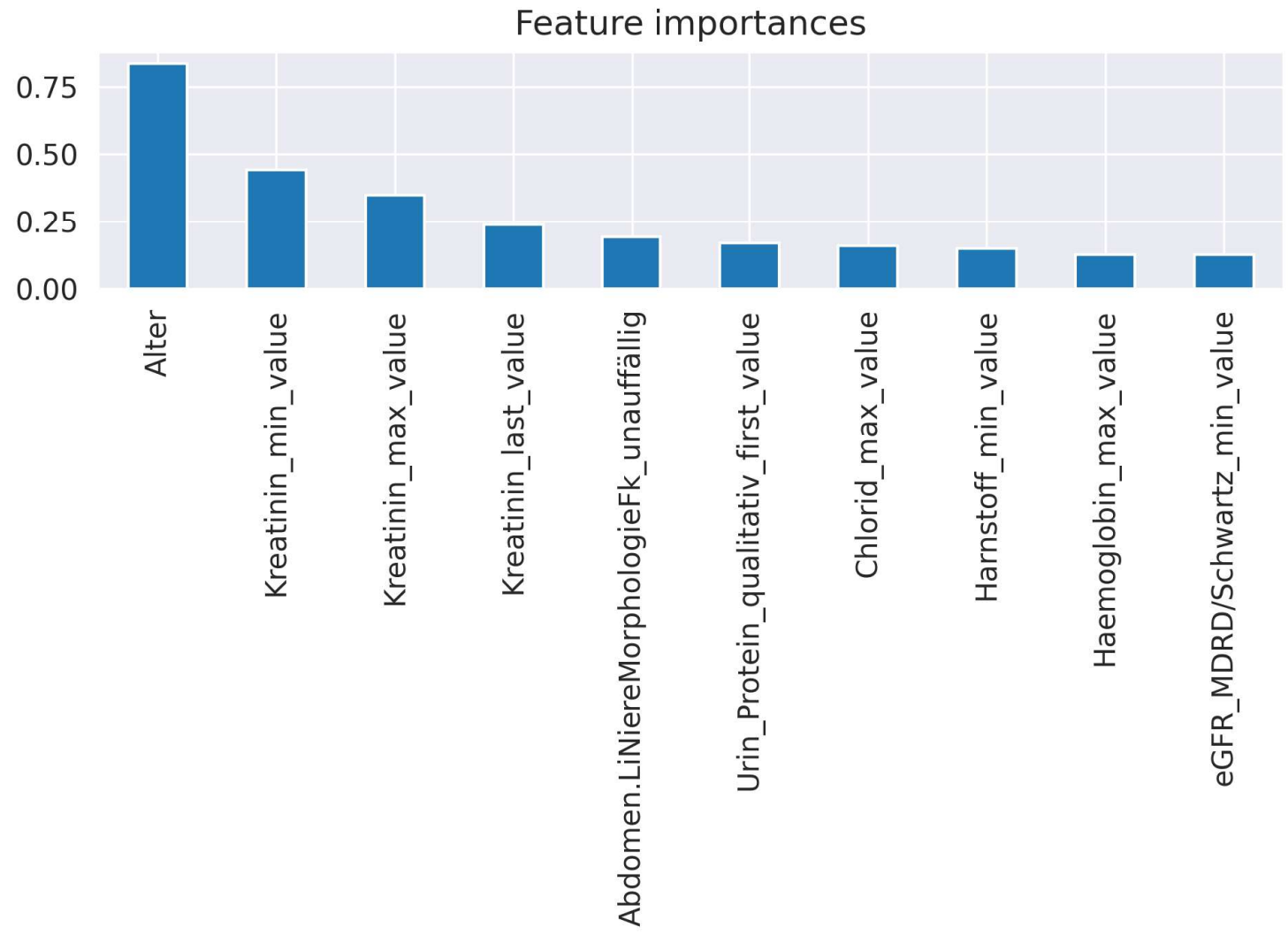

**Fig. G2** Absolute coefficients of LR for the 10 features with the largest values. As all features were z-standardized, the coefficients are changes in log-odds per standard deviation.

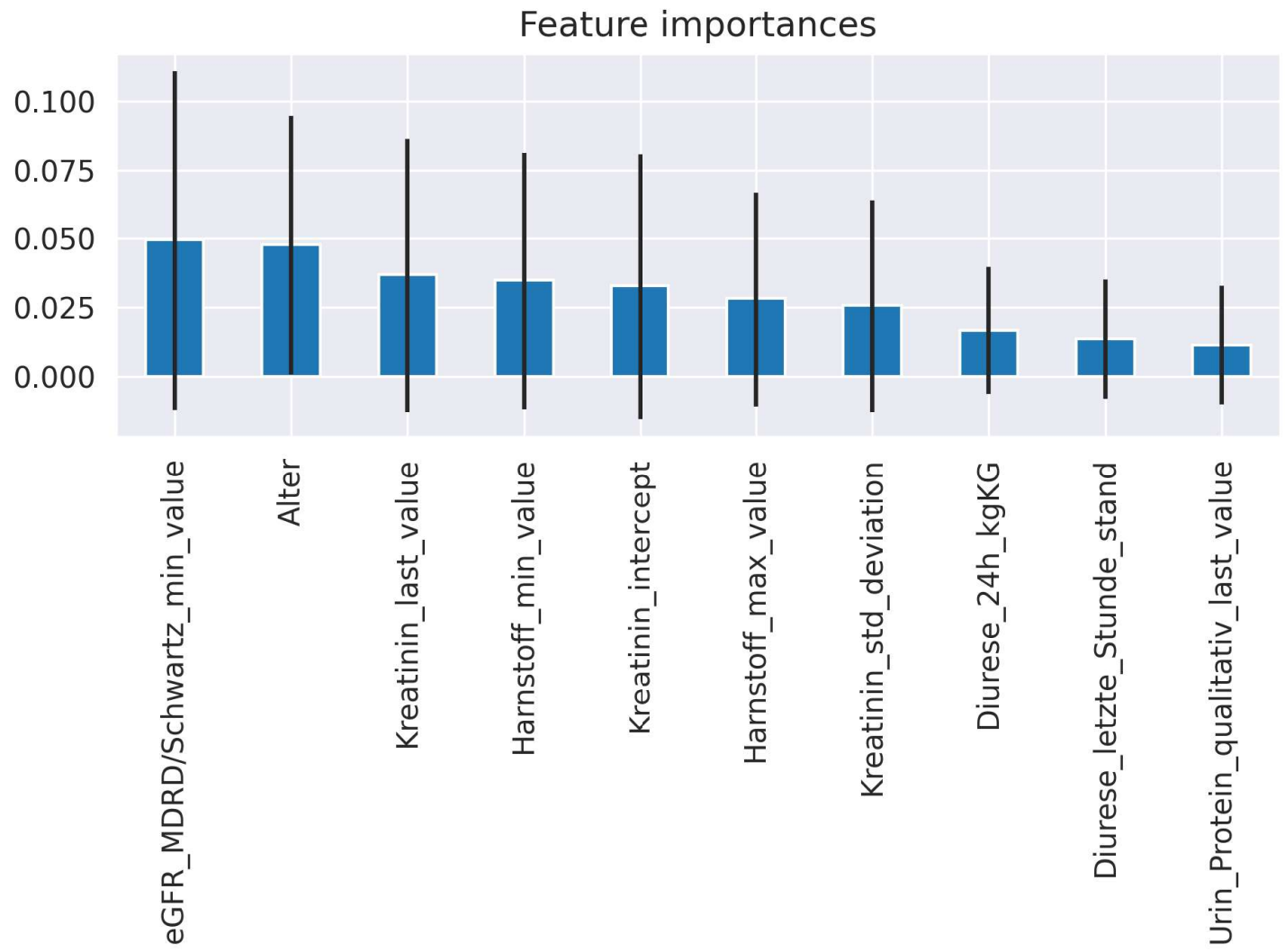

**Fig. G3** Model specific feature importance of RF (mean decrease in impurity) for the 10 most important features. Error bars show the standard deviation across trees.

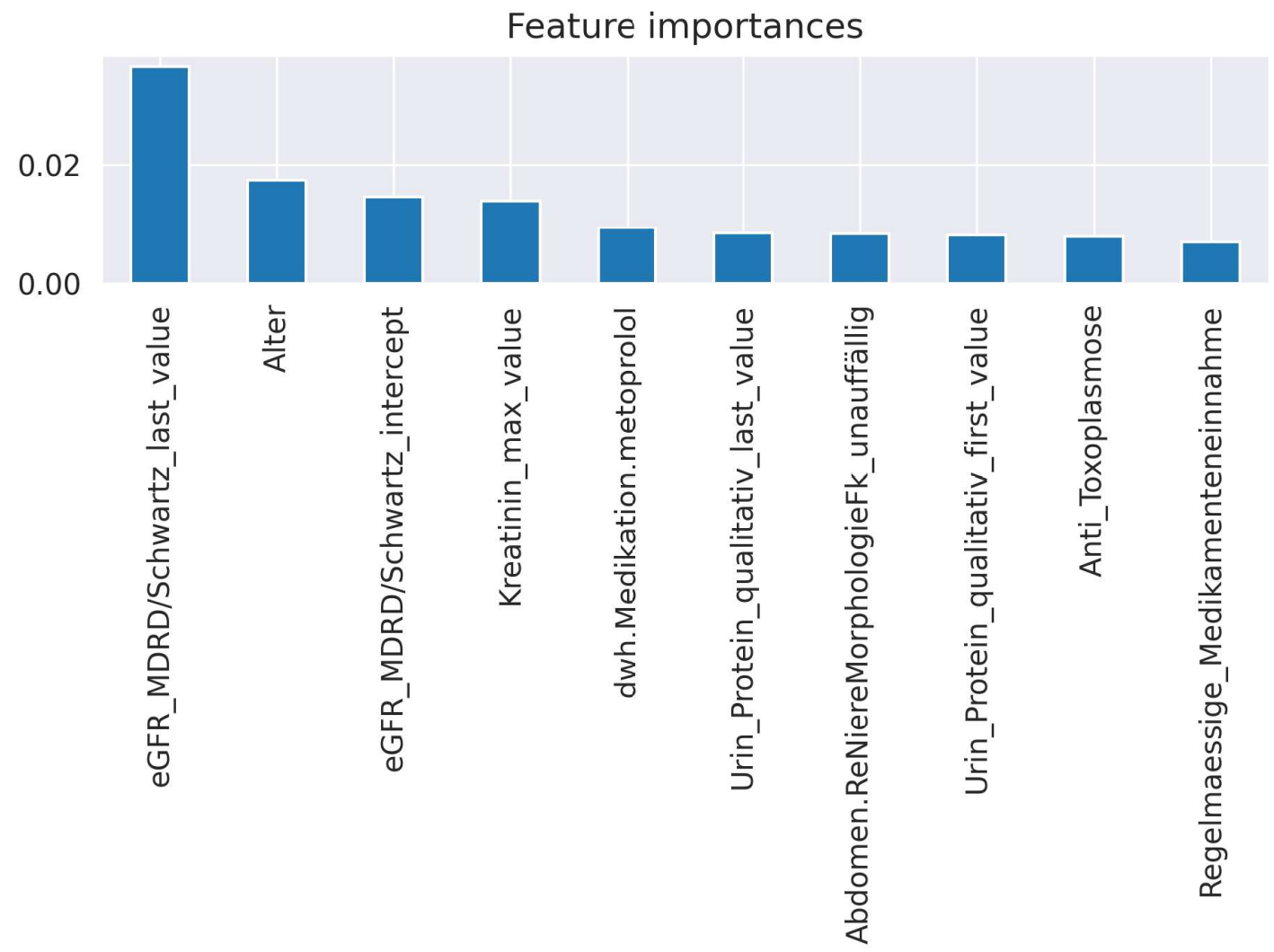

**Fig. G4** Model specific feature importance of XGB (gain) for the 10 most important features.

# TRIPOD+AI Checklist Report

Paper: Predicting Donor Kidney Discard Using Machine Learning: Model Development and Comparison Study

Generated from https://afcai.org/wp-content/uploads/2026/08/tripod-ai-checklist-1.html on 14 Nov 2026 12:07

TRIPOD 2015 items only: 100% (13 of 13) items rated; 9 Reported, 0 Not Reported, 4 N/A, 0 Unrated

New or modified items only: 100% (39 of 39) items rated; 27 Reported, 0 Not Reported, 12 N/A, 0 Unrated

Combined: 100% (52 of 52) items rated; 36 Reported, 0 Not Reported, 16 N/A, 0 Unrated

| Section | Item | Applies to | Item type | Statement | Rating | Notes |
|---|---|---|---|---|---|---|
| Title | 1 | Development; Evaluation | Carried over | Identify the study as developing or evaluating the performance of a multivariable prediction model, the target population, and the outcome to be predicted. | Reported | p. 1 |
| Abstract | 2 | Development; Evaluation | Modified | See TRIPOD+AI for Abstracts checklist. | Reported | p. 1 |
| Introduction | 3a | Development; Evaluation | Carried over | Explain the healthcare context (including whether diagnostic or prognostic) and rationale for developing or evaluating the prediction model, including references to existing models. | Reported | p. 2 |
| Introduction | 3b | Development; Evaluation | Modified | Describe the target population and the intended purpose of the prediction model in the context of the care pathway, including its intended users (eg, healthcare professionals, patients, public). | Reported | p. 3 |
| Introduction | 3c | Development; Evaluation | New | Describe any known health inequalities between sociodemographic groups. | N/A | The focus is on donor and organ characteristics, not on sociodemographic patient groups/inequalities. |
| Introduction | 4 | Development; Evaluation | Carried over | Specify the study objectives, including whether the study describes the development or validation of a prediction model (or both). | Reported | p. 3 |
| Methods — Data | 5a | Development; Evaluation | Modified | Describe the sources of data separately for the development and evaluation datasets (eg, randomised trial, cohort, routine care or registry data), the rationale for using these data, and representativeness of the data. | Reported | p. 6 |
| Methods — Data | 5b | Development; Evaluation | Carried over | Specify the dates of the collected participant data, including start and end of participant accrual; and, if applicable, end of follow-up. | Reported | p. 6 |
| Methods — Participants | 6a | Development; Evaluation | Carried over | Specify key elements of the study setting (eg, primary care, secondary care, general population) including the number and location of centres. | Reported | p. 6-7 |
| Methods — Participants | 6b | Development; Evaluation | Carried over | Describe the eligibility criteria for study participants. | Reported | p. 6-7 |
| Methods — Participants | 6c | Development; Evaluation | Modified | Give details of any treatments received, and how they were handled during model development or evaluation, if relevant. | N/A | The observations of this study are organs and not patients. |
| Methods — Data Preparation | 7 | Development; Evaluation | New | Describe any data pre-processing and quality checking, including whether this was similar across relevant sociodemographic groups. | Reported | p. 9 |
| Methods — Outcome | 8a | Development; Evaluation | Modified | Clearly define the outcome that is being predicted and the time horizon, including how and when assessed, the rationale for choosing this outcome, and whether the method of outcome assessment is consistent across sociodemographic groups. | Reported | p. 3 |
| Methods — Outcome | 8b | Development; Evaluation | New | If outcome assessment requires subjective interpretation, describe the qualifications and demographic characteristics of the outcome assessors. | N/A | Clinical endpoint: organ transplanted vs. discarded. |
| Methods — Outcome | 8c | Development; | Carried over | Report any actions to blind assessment of the outcome to be predicted. | N/A | The outcome was objectively recorded in |

| Section | Item | Applies to | Item type | Statement | Rating | Notes |
|---|---|---|---|---|---|---|
| | | Evaluation | | | | retrospective data (transplanted vs. discarded status). |
| Methods — Predictors | 9a | Development | Modified | Describe the choice of initial predictors (eg, literature, previous models, all available predictors) and any pre-selection of predictors before model building. | Reported | p. 13-15 |
| Methods — Predictors | 9b | Development; Evaluation | Modified | Clearly define all predictors, including how and when they were measured (and any actions to blind assessment of predictors for the outcome and other predictors). | Reported | p. 7, Table 1 |
| Methods — Predictors | 9c | Development; Evaluation | New | If predictor measurement requires subjective interpretation, describe the qualifications and demographic characteristics of the predictor assessors. | N/A | Predictors rely on objective data (laboratory values, vital signs, medications, ICD-10 codes) without subjective clinical evaluations. |
| Methods — Sample Size | 10 | Development; Evaluation | Modified | Explain how the study size was arrived at (separately for development and evaluation), and justify that the study size was sufficient to answer the research question. Include details of any sample size calculation. | Reported | p. 6-7 |
| Methods — Missing Data | 11 | Development; Evaluation | Modified | Describe how missing data were handled. Provide reasons for omitting any data. | Reported | p. 10-11 |
| Methods — Analytical Methods | 12a | Development | Modified | Describe how the data were used (eg, for development and evaluation of model performance) in the analysis, including whether the data were partitioned, considering any sample size requirements. | Reported | p. 9 |
| Methods — Analytical Methods | 12b | Development | Modified | Depending on the type of model, describe how predictors were handled in the analyses (functional form, rescaling, transformation, or any standardisation). | Reported | p. 17 |
| Methods — Analytical Methods | 12c | Development | Modified | Specify the type of model, rationale, all model building steps, including any hyperparameter tuning, and method for internal validation. | Reported | p. 11-16 |
| Methods — Analytical Methods | 12d | Development; Evaluation | New | Describe if and how any heterogeneity in estimates of model parameter values and model performance was handled and quantified across clusters (eg, hospitals, countries). See TRIPOD-Cluster for additional considerations. | N/A | DSO data compiled nationwide without center analysis. |
| Methods — Analytical Methods | 12e | Development; Evaluation | Modified | Specify all measures and plots used (and their rationale) to evaluate model performance (eg, discrimination, calibration, clinical utility) and, if relevant, to compare multiple models. | Reported | p. 16 |
| Methods — Analytical Methods | 12f | Evaluation | Modified | Describe any model updating (eg, recalibration) arising from the model evaluation, either overall or for particular sociodemographic groups or settings. | Reported | p. 18 |
| Methods — Analytical Methods | 12g | Evaluation | Carried over | For model evaluation, describe how the model predictions were calculated (eg, formula, code, object, application programming interface). | Reported | p. 16-17 |
| Methods — Class Imbalance | 13 | Development; Evaluation | New | If class imbalance methods were used, state why and how this was done, and any subsequent methods to recalibrate the model or the model predictions. | N/A | No class imbalanced methods were applied. |
| Methods — Fairness | 14 | Development; Evaluation | New | Describe any approaches that were used to address model fairness and their rationale. | N/A | Recipient characteristics were deliberately excluded; no demographic subgroup fairness metrics |
| Methods — Model Output | 15 | Development | New | Specify the output of the prediction model (eg, probabilities, classification). Provide details and rationale for any classification and how the thresholds were identified. | Reported | p. 17 |
| Methods — Training vs Evaluation | 16 | Development; Evaluation | Carried over | Identify any differences between the development and evaluation data in healthcare setting, eligibility criteria, outcome, and predictors. | N/A | The development and evaluation datasets were derived from a single cohort via a random 80/20 donor-level split. Consequently, healthcare setting (DSO Germany, 2020–2024), eligibility |

| Section | Item | Applies to | Item type | Statement | Rating | Notes |
|---|---|---|---|---|---|---|
| | | | | | | criteria (deceased adult donors), outcome definition (bilateral discard), and predictor availability are identical by design across development and test sets. |
| Methods — Ethical Approval | 17 | Development; Evaluation | New | Name the institutional research board or ethics committee that approved the study and describe the participant informed consent or the ethics committee waiver of informed consent. | Reported | p. 19 |
| Open Science | 18a | Development; Evaluation | Carried over | Give the source of funding and the role of the funders for the present study. | Reported | p. 31 |
| Open Science | 18b | Development; Evaluation | New | Declare any conflicts of interest and financial disclosures for all authors. | Reported | p. 32 |
| Open Science | 18c | Development; Evaluation | New | Indicate where the study protocol can be accessed or state that a protocol was not prepared. | Reported | p. 6 |
| Open Science | 18d | Development; Evaluation | New | Provide registration information for the study, including register name and registration number, or state that the study was not registered. | Reported | p. 6 |
| Open Science | 18e | Development; Evaluation | New | Provide details of the availability of the study data. | Reported | p. 31 |
| Open Science | 18f | Development; Evaluation | New | Provide details of the availability of the analytical code. | Reported | p. 31 |
| Patient and Public Involvement | 19 | Development; Evaluation | New | Provide details of any patient and public involvement during the design, conduct, reporting, interpretation, or dissemination of the study or state no involvement. | Reported | p. 6 |
| Results — Participants | 20a | Development; Evaluation | Carried over | Describe the flow of participants through the study, including the number of participants with and without the outcome and, if applicable, a summary of the follow-up time. A diagram may be helpful. | N/A | Prediction model refers to organs not to participants. |
| Results — Participants | 20b | Development; Evaluation | Modified | Report the characteristics overall and, where applicable, for each data source or setting, including the key dates, key predictors (including demographics), treatments received, sample size, number of outcome events, follow-up time, and amount of missing data. A table may be helpful. Report any differences across key demographic groups. | N/A | Prediction model refers to organs not to participants |
| Results — Participants | 20c | Evaluation | Carried over | For model evaluation, show a comparison with the development data of the distribution of important predictors (demographics, predictors, and outcome). | N/A | Prediction model refers to organs not to participants |
| Results — Model Development | 21 | Development; Evaluation | Modified | Specify the number of participants and outcome events in each analysis (eg, for model development, hyperparameter tuning, model evaluation). | Reported | p. 17, Table 2 |
| Results — Model Specification | 22 | Development | Modified | Provide details of the full prediction model (eg, formula, code, object, application programming interface) to allow predictions in new individuals and to enable third party evaluation and implementation, including any restrictions to access or reuse (eg, freely available, proprietary). | N/A | This study focuses comparing multiple machine learning algorithms rather than developing a single, deployable clinical prediction model. |
| Results — Model Performance | 23a | Development; Evaluation | Modified | Report model performance estimates with confidence intervals, including for any key subgroups (eg, sociodemographic). Consider plots to aid presentation. | Reported | p. 22 |
| Results — Model Performance | 23b | Development; Evaluation | New | If examined, report results of any heterogeneity in model performance across clusters. See TRIPOD-Cluster for additional details. | N/A | DSO data compiled nationwide without center analysis. |
| Results — Model Updating | 24 | Evaluation | Carried over | Report the results from any model updating, including the updated model and subsequent performance. | Reported | p. 24-25 |

| Section | Item | Applies to | Item type | Statement | Rating | Notes |
|---|---|---|---|---|---|---|
| Discussion | 25 | Development; Evaluation | Modified | Give an overall interpretation of the main results, including issues of fairness in the context of the objectives and previous studies. | Reported | p. 28-29 |
| Discussion | 26 | Development; Evaluation | Modified | Discuss any limitations of the study (such as a non-representative sample, sample size, overfitting, missing data) and their effects on any biases, statistical uncertainty, and generalisability. | Reported | p. 29-30 |
| Discussion — Usability in Current Care | 27a | Development | New | Describe how poor quality or unavailable input data (eg, predictor values) should be assessed and handled when implementing the prediction model. | N/A | This study was designed as a methodological model comparison rather than a prospective clinical implementation. |
| Discussion — Usability in Current Care | 27b | Development | New | Specify whether users will be required to interact in the handling of the input data or use of the model, and what level of expertise is required of users. | N/A | This study was designed as a methodological model comparison rather than a prospective clinical implementation. |
| Discussion — Usability in Current Care | 27c | Development; Evaluation | Modified | Discuss any next steps for future research, with a specific view to applicability and generalisability of the model. | Reported | p. 30 |